\documentclass[12pt]{article}

\usepackage[subpreambles=false]{standalone}

\usepackage[a4paper, margin=2cm]{geometry}
\usepackage[version=3]{mhchem} % Formula subscripts using \ce{}
\usepackage[square, numbers, sort&compress]{natbib}
\usepackage{booktabs}
\usepackage{hyperref}
\usepackage{authblk}
\usepackage{rotating} % to rotate table
\usepackage{caption}
\usepackage{placeins}
\usepackage{listings}
\usepackage{multirow}
\usepackage[figurename=Fig.]{caption}
\usepackage[capitalise]{cleveref}
\def\NCSA{National Center for Supercomputing Applications, University of Illinois Urbana-Champaign, Urbana, 61801, IL, USA}
\def\BI{Beckman Institute for Advanced Science and Technology, University of Illinois Urbana-Champaign, Urbana, 61801, IL, USA}
\def\AE{Department of Aerospace Engineering, Grainger College of Engineering, University of Illinois Urbana-Champaign, Urbana, 61801, IL, USA}
\def\MSE{Department of Mechanical Science and Engineering, Grainger College of Engineering, University of Illinois Urbana-Champaign, Urbana, 61801, IL, USA}
\def\CEE{Department of Civil and Environmental Engineering, Grainger College of Engineering, University of Illinois Urbana-Champaign, Urbana, 61801, IL, USA}
\def\KSU{Department of Industrial and Manufacturing Systems Engineering, Kansas State University, Manhattan, 66506, KS, USA}
\def\NYU{Civil and Urban Engineering Department, New York University Abu Dhabi, United Arab Emirates}

\title{Toward Signed Distance Function based Metamaterial Design: Neural Operator
Transformer for Forward Prediction and Diffusion Model for Inverse Design}
\date{}
\author%
[1,6,*]{Qibang Liu}
\author%
[1,4]{Seid Koric}
\author%
[1,7]{Diab Abueidda}
\author%
[5]{Hadi Meidani}
\author%
[2,3,**]{Philippe Geubelle}

\affil[1]{\NCSA}
\affil[2]{\BI}
\affil[3]{\AE}
\affil[4]{\MSE}
\affil[5]{\CEE}
\affil[6]{\KSU}
\affil[7]{\NYU}
\affil[*]{Corresponding author: Q. Liu, qibang@illinois.edu}
\affil[**]{Corresponding author: P. Geubelle, geubelle@illinois.edu}

\begin{document}

  \maketitle

  \begin{abstract}
    The inverse design of metamaterial architectures presents a significant challenge,
    particularly for nonlinear mechanical properties involving large
    deformations, buckling, contact, and plasticity. Traditional methods, such as
    gradient-based optimization, and recent generative deep-learning approaches
    often rely on binary pixel-based representations, which introduce jagged edges
    that hinder finite element (FE) simulations and 3D printing. To overcome
    these challenges, we propose an inverse design framework that utilizes a signed
    distance function (SDF) representation combined with a conditional diffusion
    model. The SDF provides a smooth boundary representation, eliminating the need
    for post-processing and ensuring compatibility with FE simulations and manufacturing
    methods. A classifier-free guided diffusion model is trained to generate
    SDFs conditioned on target macroscopic stress-strain curves, enabling efficient
    one-shot design synthesis. To assess the mechanical response and the quality
    of the generated designs, we introduce a forward prediction model based on Neural
    Operator Transformers (NOT), which accurately predicts homogenized stress-strain
    curves and local solution fields for arbitrary geometries with irregular
    query meshes. This approach enables a closed-loop process for general metamaterial
    design, offering a pathway for the development of advanced functional
    materials.
  \end{abstract}

  \vspace{1em} % Adds some space before keywords

  \pagebreak

  \section{Introduction}
  The evolution of material science has been revolutionized by the advent of
  additive manufacturing, facilitating the synthesis of materials with unprecedented
  customized properties. Engineers and designers are no longer confined to the
  limited options presented by natural materials but now have the ability to
  explore the vast design and property spaces offered by metamaterials. These engineered
  materials are meticulously crafted to achieve mechanical properties that were once
  considered beyond reach. Commonly realized through periodic arrangements of
  small-scale structural unit cells, metamaterials provide an innovative platform
  for exploring novel mechanical behaviors. In recent years, the production of artificial
  materials with bespoke characteristics has garnered significant interest across
  various engineering disciplines, e.g. bio-inspired material \citep{hamzehei20223d},
  soft robotics \citep{wu2019symmetry,rafsanjani2019programming}, nanophotonics \citep{staude2017metamaterial,siddique2022lessons},
  automotive \citep{tak2017metamaterial} and aerospace industries \citep{ghinet2020experimental}.
  The flexibility offered by additive manufacturing processes allows for the
  creation of architected materials, or metamaterials, at diverse scales and
  sizes. The exceptional mechanical properties of these materials, such as their
  ability to bend, stretch, compress, or respond to forces, are intricately
  linked to their internal structure, typically composed of repeating geometric patterns
  made from conventional materials.

  The material constitutive laws that govern the mechanical behavior of metamaterials
  are well developed and various numerical methods have been used to predict the
  material properties based on the unit cell structure. However, the inverse design
  of the unit cell structure that can achieve a desired material property remains
  an open challenge, especially for nonlinear mechanical properties involving
  large deformations, structural buckling, frictional contact, and plasticity.
  Traditional methods rely on a trial-and-error approach or gradient-based
  topology optimization \citep{andkjaer2010topology,chatterjee2021robust}, and
  genetic algorithms \citep{huntington2014subwavelength}. These approaches are
  computationally expensive, as they often involve iterative solutions of PDEs using
  numerical methods like finite element methods (FEM). Moreover, the efficiency of
  search algorithms declines significantly as the design space grows. Although
  such computational cost for repeatedly solving PDEs can be alleviated by using
  reduced-order models \citep{brandyberry2022multiscale} or surrogate models \citep{liu2024adaptive,cai2024towards,chugh2020surrogate,kudyshev2020machine,kollmann2020deep,abueidda2020topology},
  the gradient-based optimization process may be trapped in local minima \citep{liu2025univariate}.

  To overcome these challenges, numerous probabilistic generative models have been
  proposed for ``one-shot'' design, eliminating the need for multiple iterations.
  These include conditional generative adversarial networks (cGAN) \citep{liu2018generative,so2019designing},
  variational autoencoders (VAE) \citep{wang2020deep}, conditional variational autoencoders
  (cVAE) \citep{ma2019probabilistic}, univariate conditional variational autoencoder
  (UcVAE) \citep{liu2025univariate}, and conditional diffusion models \citep{bastek2023inverse,vlassis2023denoising,park2024nonlinear,wang2024diffmat}.
  By learning the underlying data distribution, these models generate diverse
  solutions within the design space based on a given design target.

  However, all the aforementioned methods, whether gradient-based optimization
  or generative deep-learning models, rely on a binary pixel-based representation
  of metamaterial architectures. This representation naturally introduces jagged
  edges (as shown in \cref{fig:S_geo_rep}(a)), which pose challenges when
  applied directly in finite element (FE) simulations, potentially causing significant
  mesh distortions that affect both convergence and accuracy \citep{wang2024diffmat}.
  Additionally, the jagged boundaries complicate fabrication via 3D printing, as
  sharp edges can lead to structural defects and material failure. To address
  these issues, a boundary-smoothing process is typically applied to the designed
  binary pixel-based geometry \citep{bastek2023inverse,wang2024diffmat,zheng2024text}.
  Furthermore, accurately representing geometry with a binary pixel-based
  approach requires high-resolution grids, significantly increasing computational
  costs for design.

  Instead of relying on a binary pixel-based approach, we propose an inverse
  design framework for metamaterial architectures based on a signed distance
  function (SDF) and a conditional diffusion model. Specifically, a SDF maps a
  coordinate $\textbf{x}$ to a scalar $\phi(\textbf{x})$, which represents the signed
  distance to the nearest point on the geometry's surface. A negative value
  indicates a point outside the geometry, while a positive value denotes a point
  inside. Consequently, the level set $\phi(\textbf{x})=0$ defines the shape's boundary
  (\cref{fig:S_geo_rep}(b)). The SDF representation is much more accurate than the
  binary pixel-based representation. This smooth boundary representation
  eliminates the need for additional boundary smoothing, making it directly applicable
  to finite element (FE) simulations and 3D printing. While the SDF
  representation has been utilized in level-set-based topology optimization
  methods \citep{yamada2011level,van2013level,wang2015multi}, these approaches
  are computationally expensive and face challenges in generating new holes or
  complex topologies—limitations that are effectively addressed by generative AI
  methods. To design the SDF, we employ a classifier-free guided diffusion model,
  conditioning the design process on a given target stress-strain curve. The
  diffusion model learns to synthesize the SDF from pure Gaussian noise, and the
  resulting SDF is then used to extract the geometry using a marching algorithm.
  Additionally, our method can design more general unit cell geometries by incorporating
  Gaussian random fields with the periodic Fourier method to generate the geometry
  data. This approach contrasts with previous works \citep{bastek2023inverse,park2024nonlinear},
  which generate one-quarter of the unit cell and mirror it twice to obtain the
  full unit cell, thus limiting the designs to bi-axis symmetrical geometries.

  As generative models are inherently stochastic, they can generate multiple
  solutions for a single target. This one-to-many design capability is particularly
  useful for exploring the design space and identifying optimal solutions.
  However, it is impractical to run FE simulations for each generated geometry to
  evaluate their mechanical properties and material response. Instead, a forward
  surrogate model is a prevalent choice to replace the repeated FE simulations.

  Neural operators have shown promising results in mapping infinite-dimensional input
  functions (e.g., arbitrary geometries) to output functions on query points (e.g.,
  stress-strain curves, solution fields). Notably, the Fourier Neural Operator (FNO)
  \citep{li2020fourier} and the Deep Operator Network (DeepONet) \citep{lu2021learning}
  have been effective in this domain. The Fourier Neural Operator is designed for
  fixed, regular meshes and requires a full mesh graph as input. While DeepONet can
  predict the solution field at arbitrary query points, training DeepONet
  requires that the number and location of query points remain fixed across
  samples. For arbitrary geometries, the query points typically come from
  irregular meshes, and their number varies across samples. Training DeepONet in
  such cases is challenging and often requires resampling the query points to a
  fixed number, as demonstrated in Geom-DeepONet \citep{he2024geom}. However,
  Geom-DeepONet is restricted to parametric geometries (e.g., length, thickness,
  radius, etc.) and does not generalize to fully arbitrary shapes. Moreover, it
  requires both the geometric parameters (e.g., length, thickness, radius, etc.)
  and the SDF as inputs. Although recent works like Geo-FNO \citep{li2023fourier}
  and geometry-informed FNO (GI-FNO) \citep{li2023GIFNO} extend to arbitrary geometries,
  Geo-FNO still requires a fixed number of query points during training. Moreover,
  both GI-FNO and Geo-FNO rely on projecting an irregular mesh onto a fixed
  regular mesh and then mapping it back. Subsequently, researchers have utilized
  the attention mechanism of transformers \citep{Ashish2017atten} in neural operators
  to effectively fuse information from input functions and query points \citep{li2022transformer, hao2023gnot}.
  Neural Operator Transformers (NOT) are particularly well-suited for irregular meshes
  due to the attention mechanism and allow for the prediction of solution fields
  at arbitrary query points. Unlike DeepONet, which uses a simple dot product to
  fuse information from input functions and query points, the attention
  mechanism in NOT guides each query point to focus on the relevant information
  from the input functions. This allows NOT to generalize and capture more complex
  relationships between input functions and query points.

  In this work, we propose forward NOT models to predict the macroscopic stress-strain
  curve and local solution fields for arbitrary geometries generated by the
  inverse diffusion model, given a target stress-strain curve. For stress-strain
  curve prediction, the query points correspond to fixed strain steps, whereas
  for solution field prediction, the query points consist of irregular mesh nodes.
  To handle varying numbers and locations of query points across different
  geometries, NOT is trained on batches of geometries by padding the query points
  to the maximum count within each batch, analogous to managing variable
  sequence lengths in natural language processing (NLP).

  This manuscript summarizes the key concepts, methodologies, and results of the
  closed-loop process for inverse metamaterial design and forward prediction.

  \section{Results}
  \label{results}
  \subsection{Data generation}

  \begin{figure}[h]
    \centering
    \includegraphics[width=\textwidth]{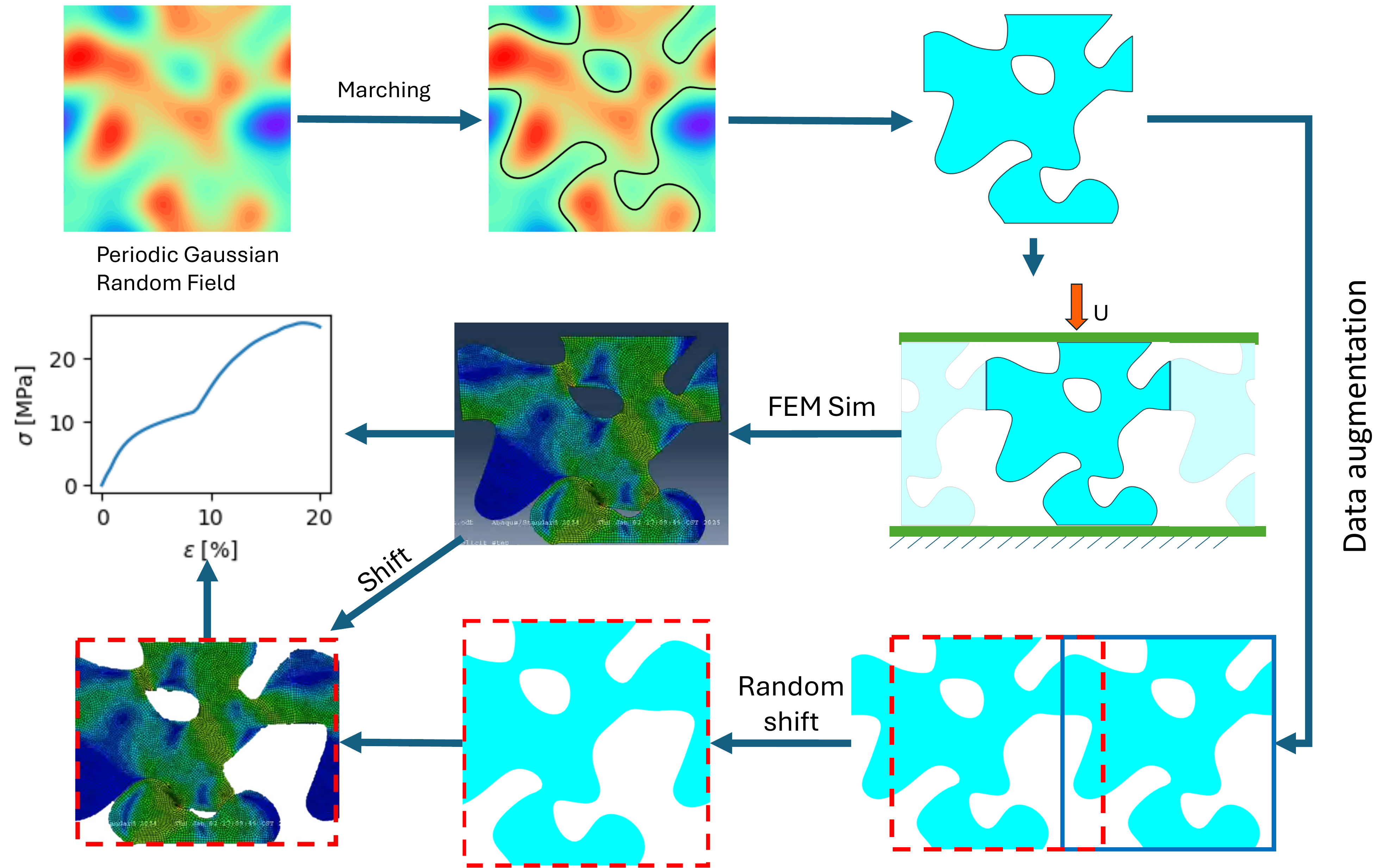}
    \caption{Workflow for data generation. Initially, a 2D periodic Gaussian
    random field is generated. The Marching algorithm is then used to extract the
    contour at a random threshold value, defining the boundary of the 2D
    periodic unit cell, with field values below the threshold set as void.
    Periodic boundary conditions and compression are applied to the unit cell
    for Abaqus simulation to obtain the macroscopic stress-strain curve. To augment
    the data, the unit cell is randomly shifted multiple times, creating different
    geometries with the same stress-strain curve. The corresponding stress and
    displacement solution fields are also shifted for the new geometries, allowing
    for data augmentation without additional FE simulations.}
    \label{fig:dataset}
  \end{figure}

  The data-driven machine learning approach requires a large dataset of paired
  training data. \cref{fig:dataset} provides an overview of the data generation process
  adopted in this work. To generate a large dataset of paired geometry, macroscopic
  stress-strain curves, and local solution fields, we first generate a 2D periodic
  Gaussian random field (GRF) with $64 \times 64$ pixels on a square domain,
  incorporating the periodic Fourier method. The GRF is then thresholded at a random
  value to extract the contour using a marching algorithm, where the extracted
  contour defines the unit cell boundary of the metamaterial. The GRF values less
  than the threshold are set as void. Connecting the contours from the marching
  algorithm with the boundary edges of the square whose GRF values are higher
  than the threshold provides a closed manifold geometry. We also evaluate the signed
  distance function (SDF) of the geometry on a regular and uniform $120 \times 12
  0$ grid, which represents the geometry as input or output of the deep learning
  models. This method generates arbitrary, random, more general periodic unit cell
  geometries and a vast design space.

  The generated geometries are then imported into the commercial finite element code
  ABAQUS\textsuperscript{\textregistered} 2024 to compute the macroscopic stress-strain
  curve and the local solution fields. We fix the bottom of the unit cell along
  the y-direction ($u_{y}=0$) and deform the unit cell by applying a vertical
  displacement to the top edge of the unit cell with a maximum applied
  compressive strain of 0.2. Periodic boundary conditions are applied along the left
  and right sides of the unit cell. Self-contact is enabled to prevent the unit cell from collapsing during
  compression. Simulation details are provided in the methods \cref{sec:fesim}.

  \begin{figure}[htbp]
    \centering
    \includegraphics[width=\textwidth]{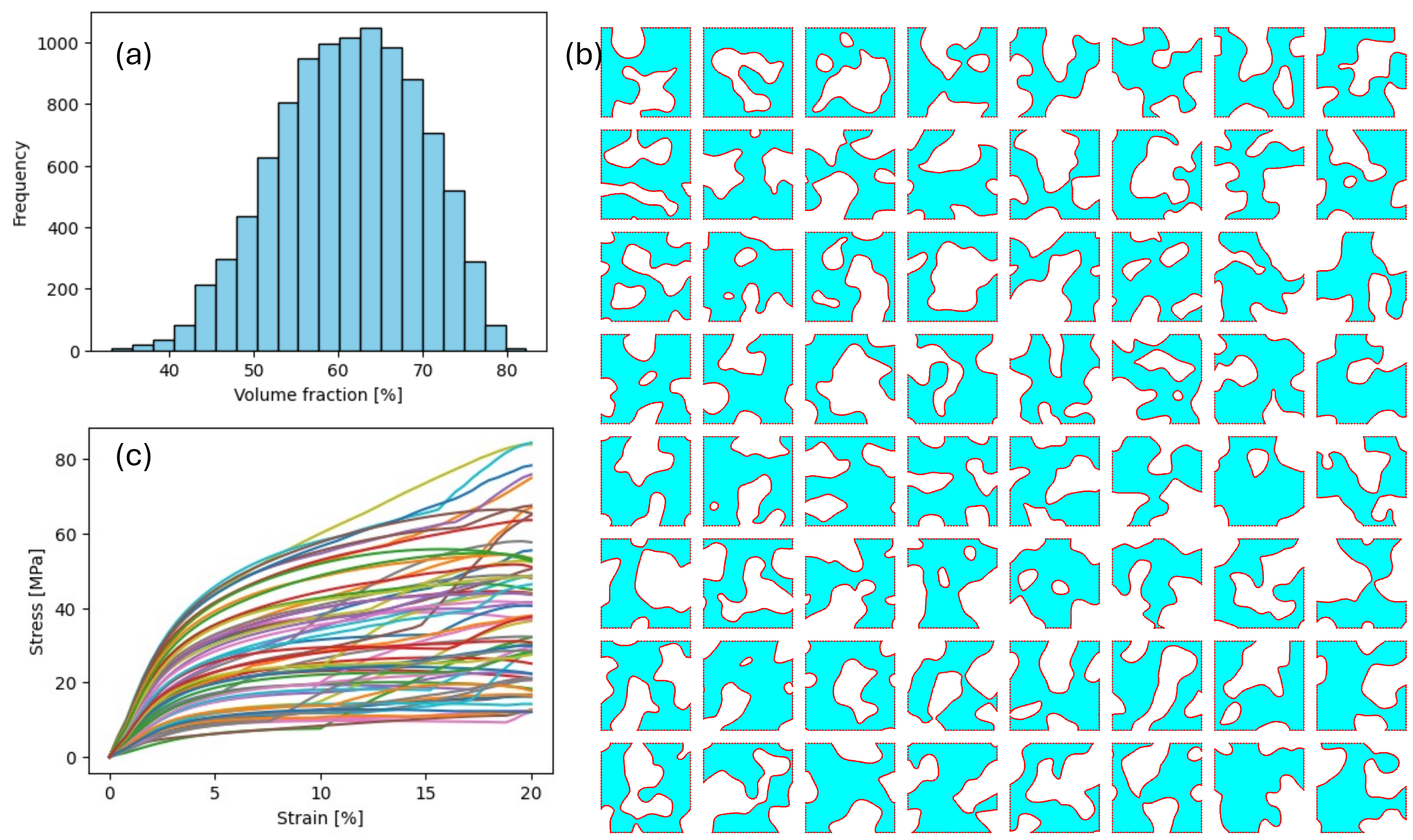}
    \caption{Illustration of training samples. (a) Volume fraction distribution of
    the dataset. (b) 64 typical microstructure geometries. (c) Corresponding stress-strain
    curves for the geometries shown in (b).}
    \label{fig:X_dataset}
  \end{figure}

  Based on this method, we randomly generate 10k geometries and their
  corresponding stress-strain curves and solution fields (e.g., Mises stress,
  displacement). \cref{fig:X_dataset} illustrates the generated dataset, where (a)
  shows the volume fraction of the 10k samples, (b) displays 64 examples of the
  generated geometries, and (c) presents their corresponding stress-strain curves.

  To alleviate overfitting, we augment the dataset by randomly shifting the unit
  cell in the x direction multiple times, as shown in \cref{fig:dataset}. This
  shifted geometry exhibits the same material response as the original geometry
  due to the periodic nature of the geometry and boundary conditions. The solution
  fields of the shifted geometry can be derived by shifting the original solution
  fields accordingly. With this data augmentation, we have about 73k samples for
  training and testing the deep learning models.

  \subsection{Forward prediction}

  \begin{figure}[!h]
    \centering
    \includegraphics[width=\textwidth]{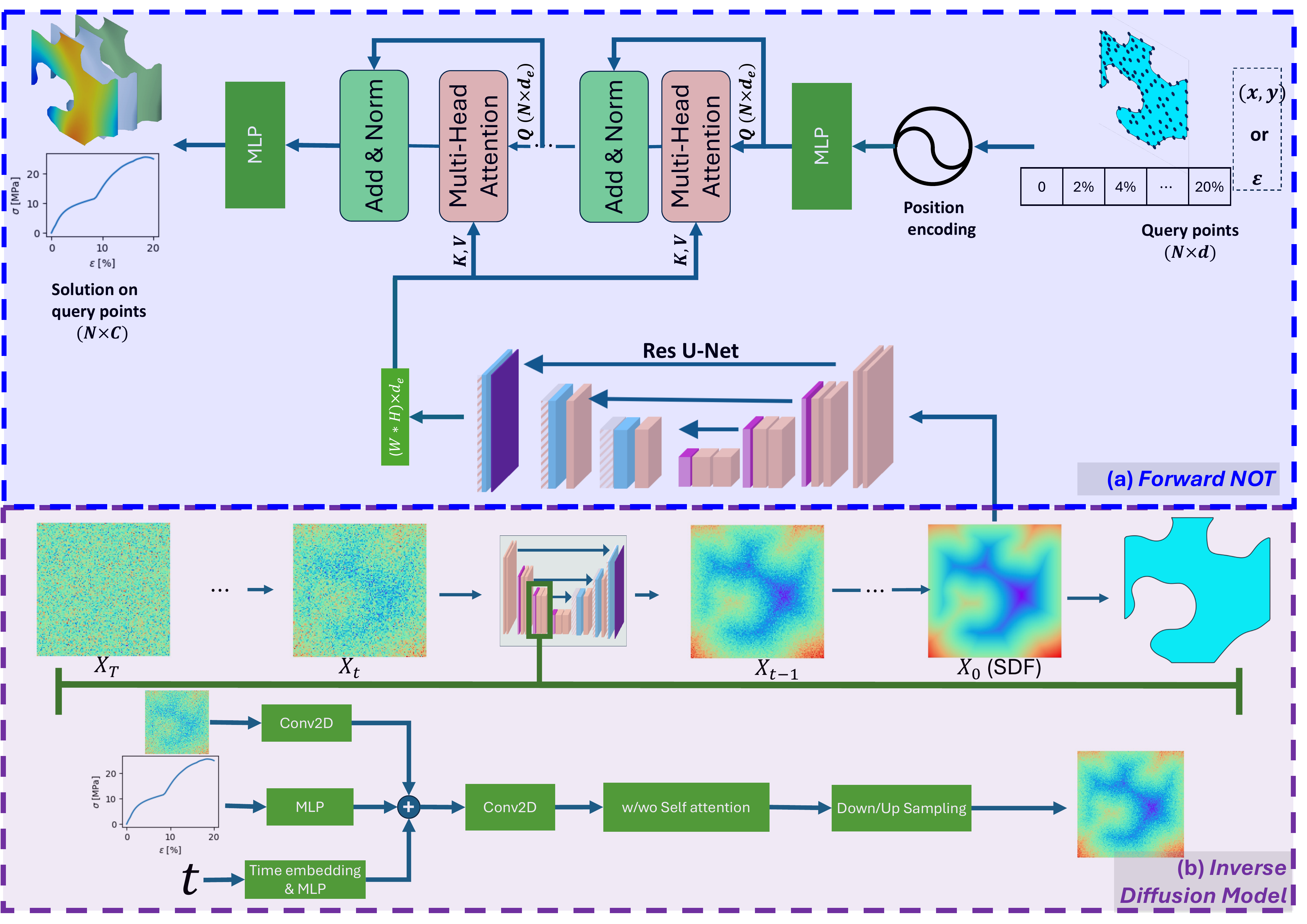}
    \caption{Overview of model architectures. The inverse denoising diffusion model
    (b) is used to design the micro-structure geometry represented by SDF, given
    the target stress-strain curve. The noise estimator is a residual U-Net architecture,
    and each residual block takes as input the image from previous layer, the time
    step $t$, and the stress-strain curve. The iterative denoising process generates
    the SDF from pure Gaussian noise, from which the geometry is extracted using
    a marching algorithm. Two forward neural operator transformers (a) are developed
    to predict the macroscopic stress-strain curve and local solution fields for
    arbitrary geometries. A residual U-net encodes the geometry represented by
    SDF into key (K) and value (V) for the attention mechanism. The query points,
    either the strain $\varepsilon$ or node coordinates $(x,y)$, are encoded using
    NeRF positional encoding \citep{mildenhall2021nerf} and a multi-layer
    perceptron (MLP) to form the query (Q) for the attention mechanism. This mechanism
    fuses the geometry information and query points information, guiding each
    query point to focus on relevant geometry information. The output of a few attention
    blocks is then decoded using an MLP to the solution fields on the query points.}
    \label{fig:models}
  \end{figure}

  \begin{figure}[htbp]
    \centering
    \includegraphics[width=\textwidth]{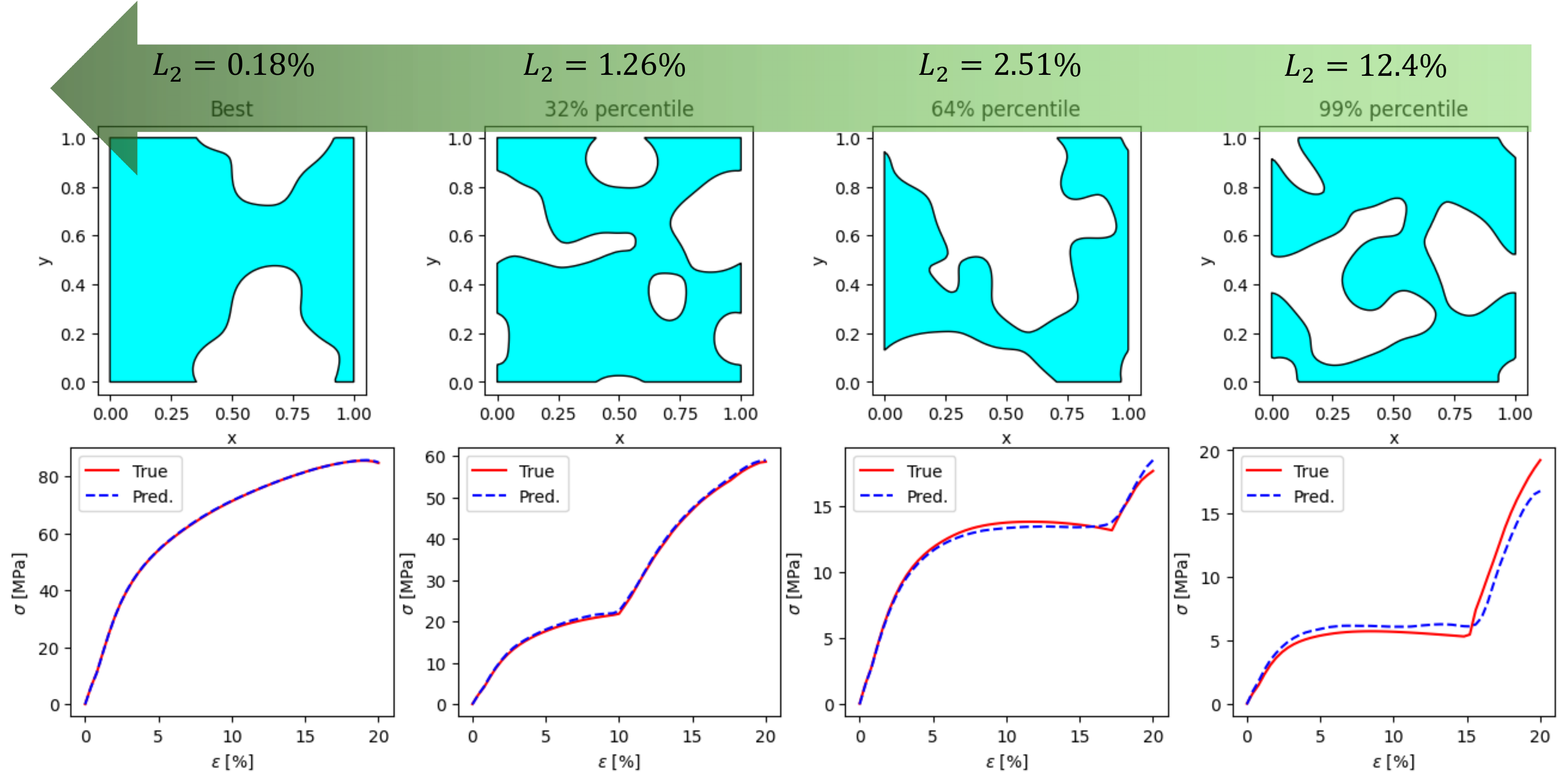}
    \caption{Comparison between the true and predicted stress-strain curves of
    the test data, arranged from best (left) to 99th percentile (right). The
    first row shows the geometries while the second row presents the
    corresponding macroscopic stress-strain curves.}
    \label{fig:not_ss_perform}
  \end{figure}

  \begin{figure}[htb]
    \centering
    \includegraphics[width=\textwidth]{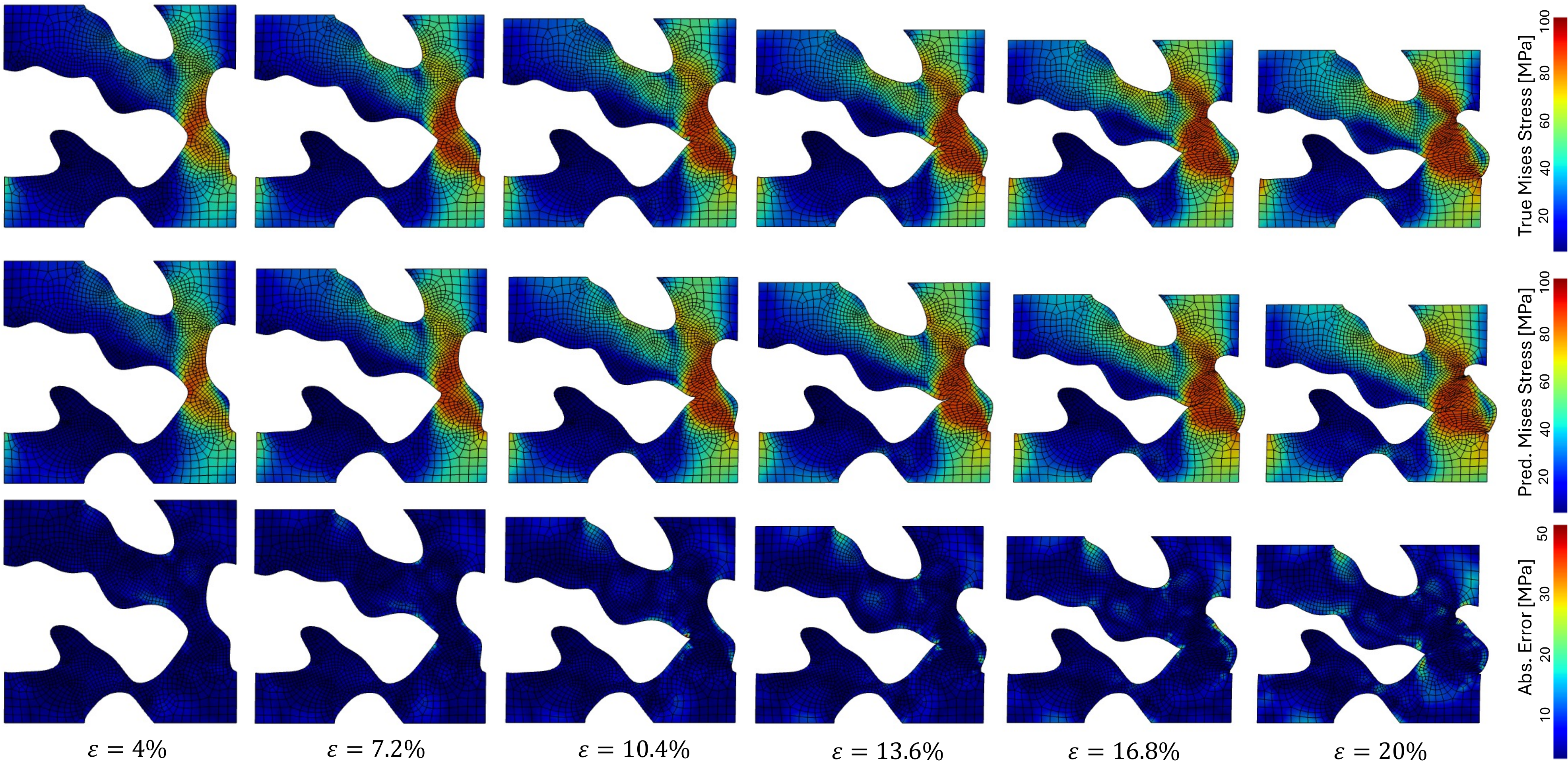}
    \caption{Comparison of the Mises stress and displacement predictions with FE
    ground truth for the median case of the test data at different strain
    $\varepsilon$ steps. The first row shows the true Mises stress under the true
    deformed shape, the second row shows the predicted Mises stress under the
    predicted deformed shape, and the third row shows the absolute error of the Mises
    stress under the true deformed shape.}
    \label{fig:not_su_perform_median}
  \end{figure}
  Neural operator transformers (NOT) map infinite-dimensional input functions to
  output functions on arbitrary query points:
  \begin{equation}
    \label{eq:NOT}G_{\theta}:\mathcal{F}\rightarrow \mathcal{G}(\textbf{x}),
  \end{equation}
  where ${G}_{\theta}$ is the neural operator with learnable parameter $\theta$,
  $\mathcal{F}$ is the space of input functions (e.g., arbitrary geometries), and
  $\mathcal{G}$ is the space of output functions (e.g., stress-strain curves, solution
  fields) on query points $\textbf{x}$.

  The attention mechanism \citep{Ashish2017atten} in transformers allows the model
  to fuse information from the input functions and the query points, guiding
  each query point to focus on the relevant information from the input functions.
  It can be described as:
  \begin{equation}
    \text{Attention}(Q,K,V)=\text{softmax}\left(\frac{QK^{T}}{\sqrt{d_{e}}}\right
    )V,
  \end{equation}
  where $Q \in \textbf{R}^{n \times d_{e}}$ is the Query matrix, and $K, V \in \textbf
  {R}^{n_k \times d_{e}}$ are the Key and Value matrices, $d_{e}$ is the
  embedding dimension, $n$ is the number of query points, and $n_{k}$ is the
  length of the key and value.

  We developed two neural operator transformers for forward prediction solutions
  from the unit cell geometry represented by SDF, as shown in \cref{fig:models}(a).
  The first model predicts the stress-strain curve, where the query points are
  the strain steps. The second model predicts the solution fields (e.g., Mises
  stress, displacement) at 26 strain steps, with query points being the irregular
  mesh nodes. Since the SDF is defined on a regular and uniform grid, we use a residual
  U-Net architecture to encode the geometry into key (K) and value (V) for the attention
  mechanism. The query points, either the strain $\varepsilon$ or node coordinates
  $(x,y)$, are encoded using NeRF (Neural Radiance Fields) positional encoding
  \citep{mildenhall2021nerf} and MLP to the Query (Q) for the attention mechanism.
  To stabilize training, residual connections and layer normalization are applied
  in each attention block. The output of a few attention blocks is decoded to
  the solution fields on the query points. We provide details on the methods and
  implementations in the Method \cref{sec:methods} and the supplementary \cref{s_sec:forward}.

  The performance of the forward model for predicting the stress-strain curve from
  arbitrary unit cell geometries is illustrated in \cref{fig:not_ss_perform},
  which shows a comparison between the true and predicted stress-strain curves
  of the test data, arranged in terms of $L_{2}$ relative error from best (left)
  to the 99th percentile (right). Further performance details for predicting the
  stress-strain curve are presented in Supplementary Information (\cref{fig:S_notss_perform}),
  which shows the MSE loss of training history and the $L_{2}$ relative error distribution
  of the test data. The overall mean $L_{2}$ relative error of the test data is
  2.6\% with a standard deviation of 2.4\%, indicating the model's high accuracy
  in predicting the stress-strain curve.

  We also demonstrate in \cref{fig:not_su_perform_median} the performance of the
  forward model in predicting the Mises stress and displacement fields for the median
  case of the test data. The first row displays the true (FE-predicted) Mises
  stress field plotted on the true deformed shape at different strain steps. The
  second row shows the predicted (by the NOT) Mises stress under the predicted deformed
  shape, while the third row presents the corresponding absolute error in Mises
  stress plotted on the true deformed shape of the unit cell. The $L_{2}$ relative
  error for the median case is 10.1\%. Additionally, we present the best and
  worst cases of the test data in Supplementary Information (\cref{fig:S_notsu_perform_best}
  and \cref{fig:S_notsu_perform_worst}). The MSE loss of the training history and
  the $L_{2}$ relative error distribution of the test data for Mises stress and displacement
  solutions are shown in Supplementary Information \cref{fig:S_notsu_perform}. The overall mean $L_{2}$
  relative error of the test data is 10.3\% with a standard deviation of 4.6\%, indicating
  the model's high accuracy in predicting the solution fields for arbitrary geometries.

  \subsection{Inverse design by diffusion model}
  In this section, we first describe the classifier-free guided diffusion model \citep{ho2022classifier}
  trained to design the micro-structure of the unit cell of the metamaterial that
  can achieve a target stress-strain curve. The label data $y$ is the stress-strain
  curve obtained from the FE simulation of the corresponding geometry, represented
  by SDF ($x_{0}$).

  Given a sample $\mathbf{x}_{0}$ from a prior data distribution $\mathbf{x}_{0}\sim
  q(\mathbf{x})$, the forward diffusion process of the denoising diffusion
  probabilistic model adds a small amount of Gaussian noise to the sample over
  $T$ steps. This results in a sequence of samples $\mathbf{x}_{0}, \mathbf{x}_{1}
  , \ldots, \mathbf{x}_{T}$. Each step in the diffusion process is controlled by
  a Gaussian distribution,
  \begin{equation}
    \label{eq:diffusing}q\left(\mathbf{x}_{t}|\mathbf{x}_{t-1}\right)=\mathcal{N}
    \left(\mathbf{x}_{t}; \sqrt{1-\beta_{t}}\mathbf{x}_{t-1}, ~\beta_{t}\mathbf{I}
    \right),\left\{\beta_{t}\in(0,1)\right \}_{t=1}^{T}.
  \end{equation}

  This diffusion process has the elegant property that we can sample
  $\mathbf{x}_{t}$ at any time step $t$ using the reparametrization trick,
  \begin{equation}
    \label{eq:sampling}\mathbf{x}_{t}= \sqrt{\bar{\alpha}_{t}}\mathbf{x}_{0}+ \sqrt{1
    - \bar{\alpha}_{t}}\epsilon_{t},
  \end{equation}
  where $\alpha_{t}= 1 - \beta_{t}$,
  $\bar{\alpha}_{t}= \prod_{i=1}^{t}\alpha_{i}$, and $\epsilon_{t}\sim \mathcal{N}
  (0, \mathbf{I})$.

  The reverse diffusion process $q(\mathbf{x}_{t-1}| \mathbf{x}_{t}, \mathbf{x}_{0}
  )$ can be approximated by a posterior distribution $p_{\theta}(\mathbf{x}_{t-1}
  | \mathbf{x}_{t}) \sim \mathcal{N}\left (\boldsymbol{\mu}_{\theta}\left ( \mathbf{x}
  _{t}, t\right ), \mathbf{\Sigma}_{\theta}\left( \mathbf{x}_{t}, t\right )\right
  )$, where the variance is
  $\mathbf{\Sigma}_{\theta}\left(\mathbf{x}_{t}, t\right) = \frac{1 - \bar{\alpha}_{t-1}}{1
  - \bar{\alpha}_{t}}\beta_{t}\mathbf{I}$, and $\boldsymbol{\mu}_{\theta}\left(\mathbf{x}
  _{t}, t\right)$ is the predicted mean defined as
  \begin{equation}
    \label{eq:approx_mu}\boldsymbol{\mu}_{\theta}\left(\mathbf{x}_{t}, t\right )
    = \frac{1}{\sqrt{\alpha_{t}}}\left(\mathbf{x}_{t}- \frac{1 - \alpha_{t}}{\sqrt{1
    - \bar{\alpha}_{t}}}\epsilon_{\theta}\left(\mathbf{x}_{t}, t\right)\right ),
  \end{equation}
  in which $\epsilon_{\theta}\left(\mathbf{x}_{t}, t\right)$ is the noise estimator
  approximated by a neural network (NN) with learnable parameters $\theta$.

  This reverse generative process is random and not controlled by any specific target.
  In specific design tasks, we aim to generate fields given the condition of a target
  such as the stress-strain curve, which requires training the NN
  $\epsilon_{\theta}\left(\mathbf{x}_{t}, t\right)$ with conditional information.
  To incorporate the condition information $\mathbf{y}$ into the diffusion process,
  we use the classifier-free guidance method \citep{ho2022classifier}. In the
  classifier-free guided diffusion model, the unconditional noise estimators $\epsilon
  _{\theta}\left(\mathbf{x}_{t}, t, \mathbf{c}=\emptyset\right)$ and the conditional
  one $\epsilon_{\theta}\left (\mathbf{x}_{t}, t, \mathbf{c}=\mathbf{y}\right)$ of
  $p_{\theta}(\mathbf{x}|\mathbf{c})$ are trained in a single NN $\epsilon_{\theta}
  \left(\mathbf{x}_{t}, t, \mathbf{c}\right)$, in which condition information
  $\mathbf{c}$ is randomly set as $\mathbf{c}=\emptyset$ or
  $\mathbf{c}=\mathbf{y}$. During the reverse inference process,
  $\epsilon_{\theta}$ in \cref{eq:approx_mu} is replaced by the linear summation
  of conditional and unconditional noise estimators,
  \begin{equation}
    \label{eq:guidance}\epsilon_{\theta}\left(\mathbf{x}_{t}, t, \mathbf{c}\right
    ) = (1 + w)\epsilon_{\theta}\left(\mathbf{x}_{t}, t, \mathbf{c}=\mathbf{y}\right
    ) - w \epsilon_{\theta}\left(\mathbf{x}_{t}, t, \mathbf{c}=\emptyset \right),
  \end{equation}
  where $w \ge 0$ is the guidance weight. Further details of the classifier-free
  guided diffusion model are summarized in the Supplementary
  \cref{s_sec:diffusion}.

  As shown in \cref{fig:models}(b), the noise estimator is approximated by a
  residual U-Net which takes the input of the current image $x_{t}$, the time step
  $t$, and the condition information $\mathbf{c}=y$ or $\mathbf{c}=\emptyset$. We
  use a 2D convolutional layer and an MLP to extract the features of the image from
  previous level of U-Net and the target stress-strain curve $y$. The time step
  $t$ is encoded using a time embedding layer and an MLP. The outputs of these three
  branches are concatenated and fed into a 2D convolutional layer, optionally
  followed by a self-attention layer. The output is then processed with down-sampling
  or up-sampling layers for the next level of the U-Net. We provide the detailed
  methods and implementation in the Method \cref{sec:methods} and Supplementary
  \cref{s_sec:inverse}.

  \begin{figure}[htbp]
    \centering
    \includegraphics[width=\textwidth]{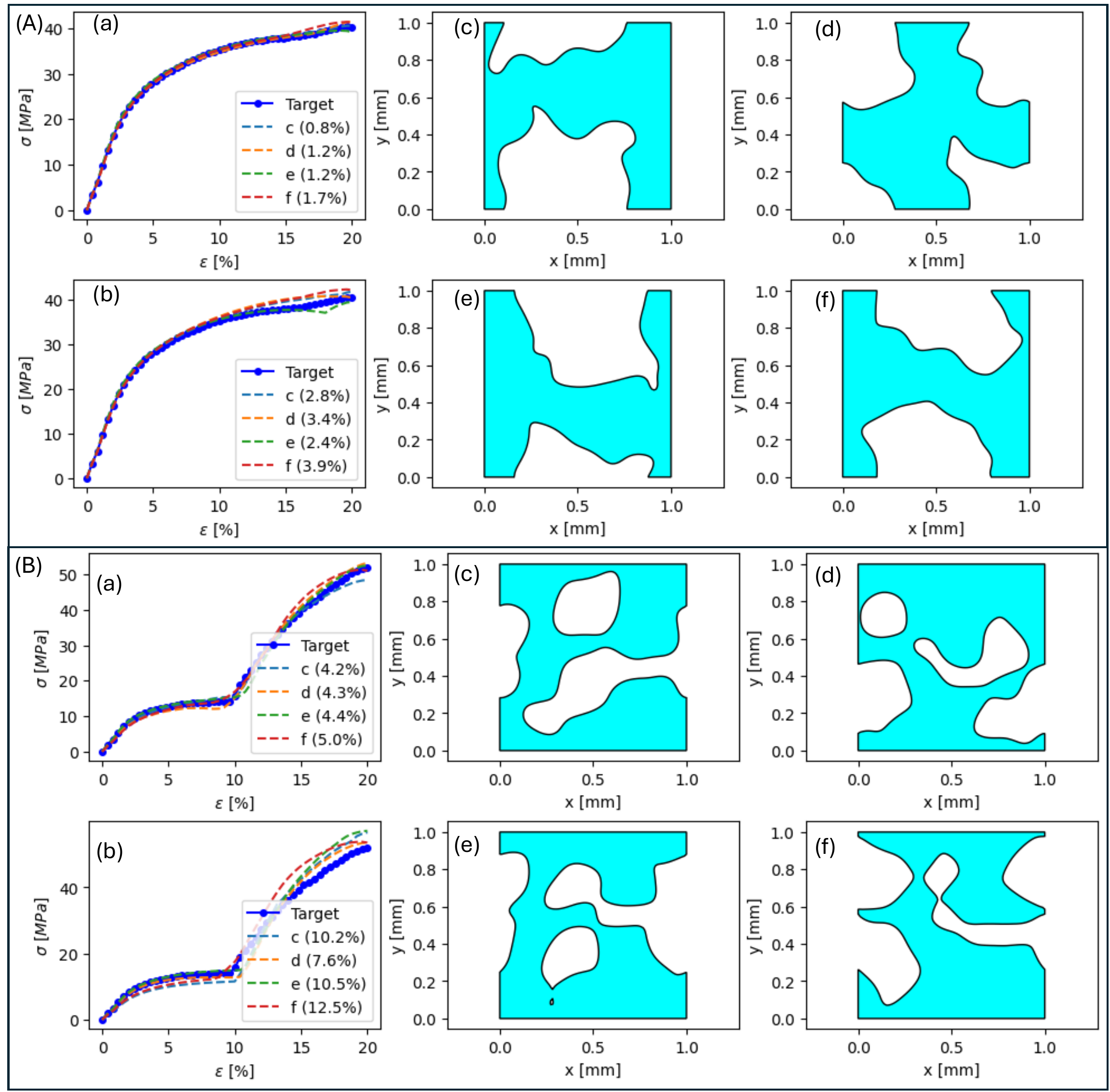}
    \caption{One-to-many unit cell design. Two targets, (A) and (B), of stress-strain
    curves are randomly selected from the test dataset and fed into the inverse diffusion
    model with guidance weight $w=10$ to generate 200 geometry solutions. These
    solutions are then evaluated using the forward model to predict the stress-strain
    curves. We select the 4 best design results based on the $L_{2}$ relative error
    between the target and the forward model predictions, shown in (c-f). (a) shows
    the comparison between the target and the predictions of (c-f) using the forward
    model, with the $L_{2}$ relative error indicated in the legend. (b) shows the
    comparison between the target and the stress-strain curves obtained using Abaqus
    simulation for (c-f).}
    \label{fig:X_test_design}
  \end{figure}

  \begin{figure}[htbp]
    \centering
    \includegraphics[width=\textwidth]{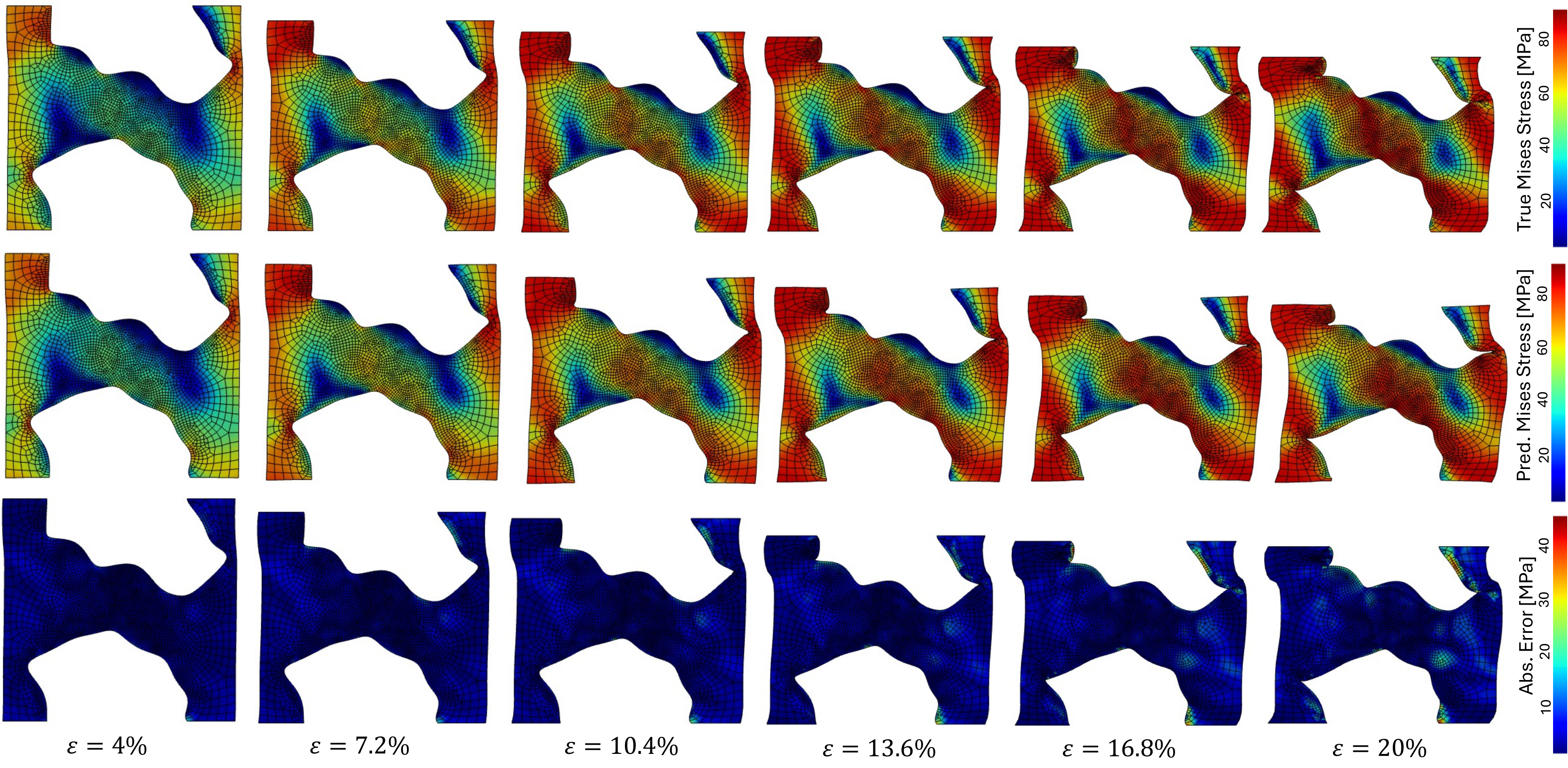}
    \caption{Comparison between the Mises stress and displacement field
    predictions with the corresponding FE ground truth at different strain
    $\varepsilon$ steps for the designed structure of \textbf{\cref{fig:X_test_design}(Af)}.
    The first row shows the true Mises stress field displayed on the true
    deformed shape; the second row shows the predicted Mises stress field onto the
    predicted deformed shape; and the third row shows the absolute error in
    Mises stress plotted on the true deformed shape.}
    \label{fig:X_test_design_SU_Af}
  \end{figure}

  \begin{figure}[htbp]
    \centering
    \includegraphics[width=\textwidth]{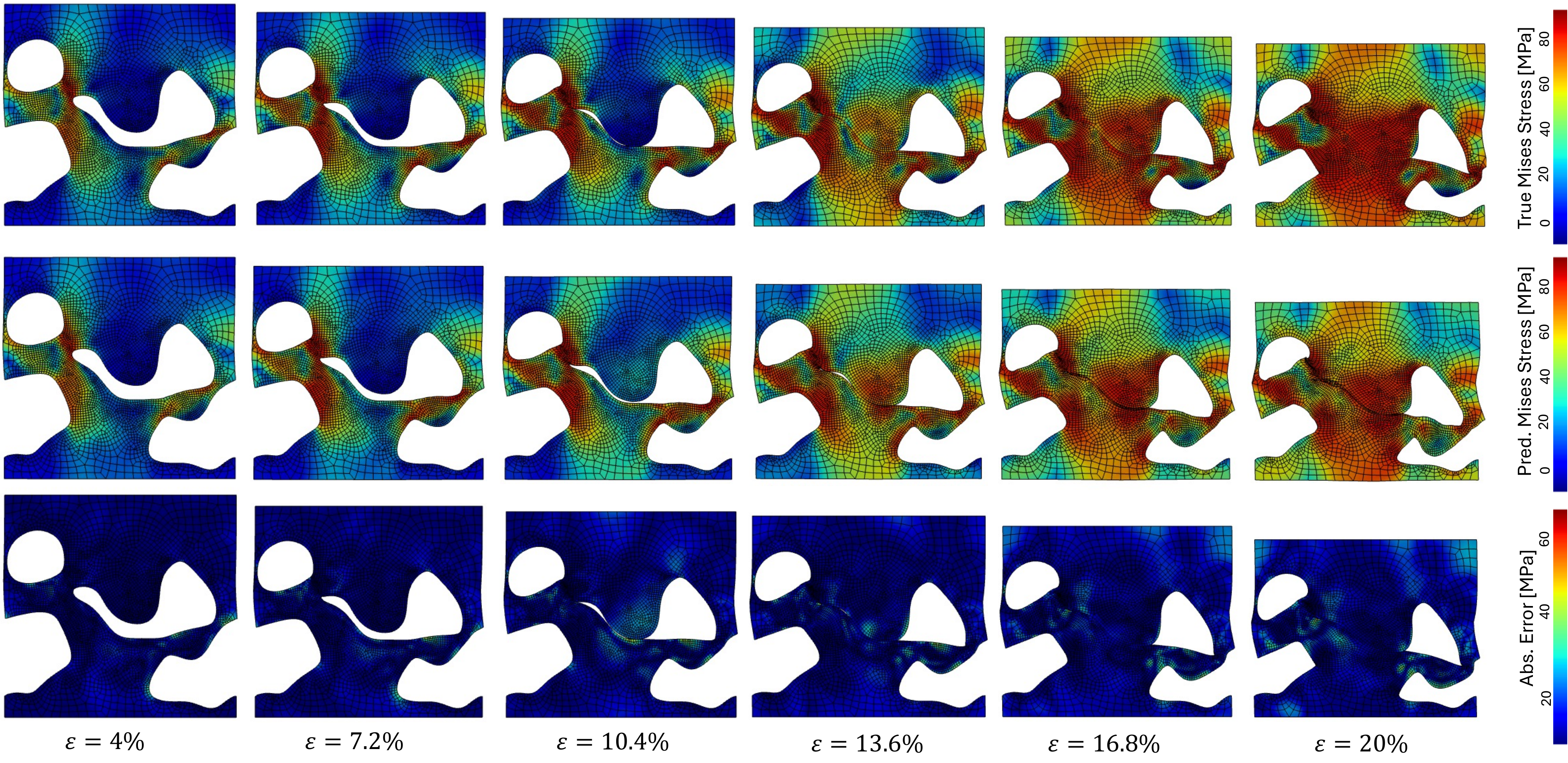}
    \caption{Comparison between the Mises stress and displacement field
    predictions with the corresponding FE ground truth at different strain
    $\varepsilon$ steps for the designed structure of \textbf{\cref{fig:X_test_design}(Bd)}.
    The first row shows the true Mises stress under the true deformed shape; the
    second row shows the predicted Mises stress under the predicted deformed shape;
    and the third row shows the absolute error of the Mises stress under the
    true deformed shape.}
    \label{fig:X_test_design_SU_Bd}
  \end{figure}

  Such a generative diffusion model can perform one-to-many designs, where
  giving a single target stress-strain curve generates multiple unit cell
  geometries that satisfy the target stress-strain curve. We demonstrate the trained
  diffusion model for such one-to-many material architecture designs by randomly
  selecting two target stress-strain curves from the test dataset and feeding them
  into the inverse diffusion model for 200 geometry solutions. The generated
  geometries are then fed into the forward model to predict the stress-strain
  curve. We select the four best design results in terms of $L_{2}$ relative error
  between the target and prediction of the forward model, as shown in
  \cref{fig:X_test_design}(c-f). The comparison between the target and
  prediction of the forward model is shown in \cref{fig:X_test_design}(a), with
  the $L_{2}$ relative error in the legend bracket. The corresponding stress-strain
  curves obtained using Abaqus simulation are shown in \cref{fig:X_test_design}(b).
  The results show that the designed geometries can achieve the target stress-strain
  curve with high accuracy. With the designed geometry, we can further predict the
  solution fields (e.g., Mises stress, displacement) at different strain steps
  using our second forward NOT. The comparison of solution fields between the true
  and predicted results for the generated geometry of \cref{fig:X_test_design}(Af) and \cref{fig:X_test_design}(Bd)
  are shown in \cref{fig:X_test_design_SU_Af} and \cref{fig:X_test_design_SU_Bd},
  respectively.

  \begin{figure}[htbp]
    \centering
    \includegraphics[width=\textwidth]{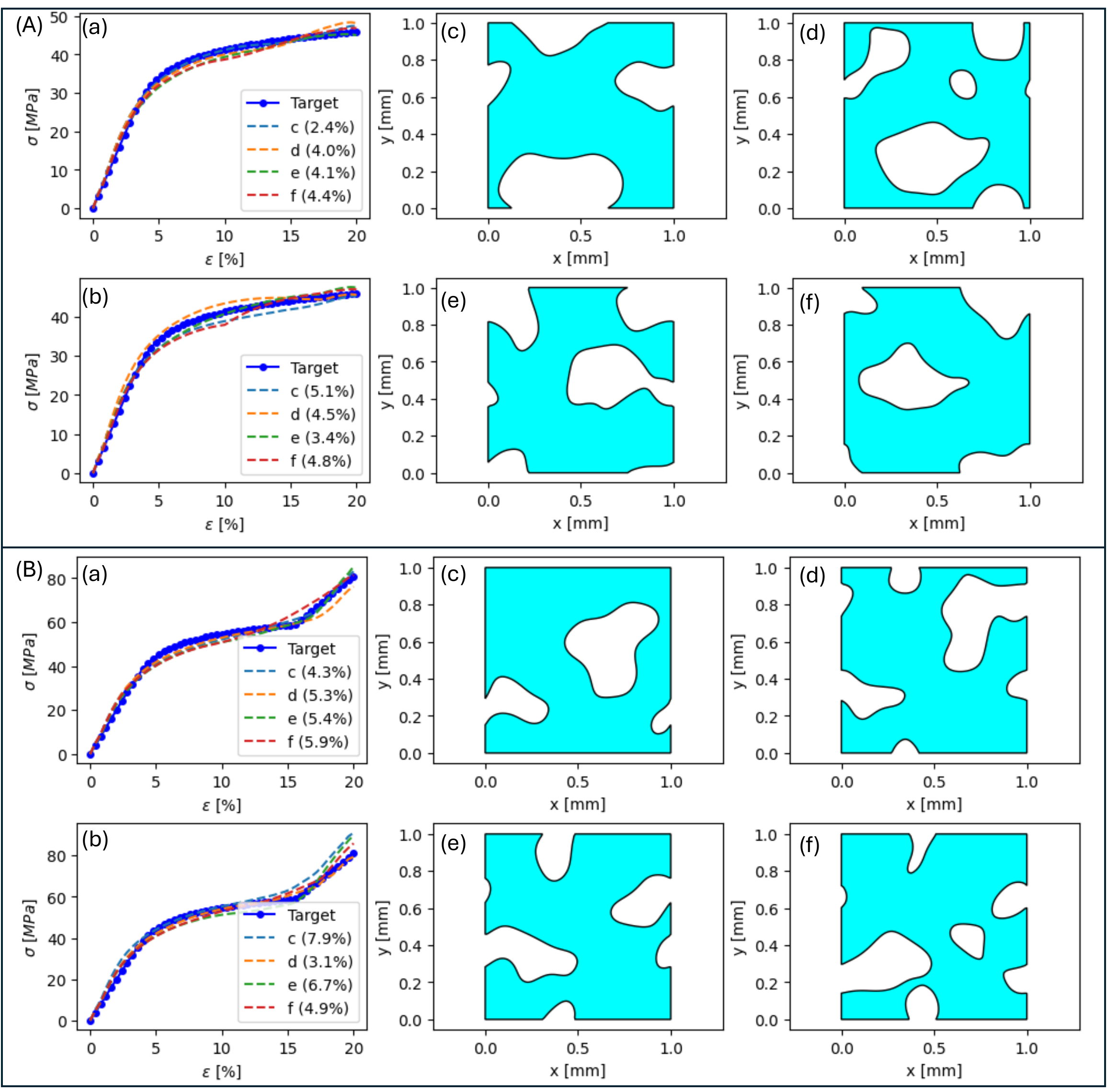}
    \caption{On-demand one-to-many unit cell design. Two targets of stress-strain
    curves are generated using the Ramberg-Osgood equation \citep{gadamchetty2016practical}.
    The top case has material properties of Young's modulus $E=800$ MPa and reference
    yield stress $\sigma_{0}=30$ MPa. The bottom case is obtained using $E=1000$
    MPa and $\sigma_{0}=40$ MPa, but stress only follows the Ramberg-Osgood equation
    for $\varepsilon \leq 0.156$. For $\varepsilon \in [0.156,0.2]$, the stress increases
    linearly with Young's modulus of $500$ MPa. These two target curves are fed
    into the inverse diffusion model for 500 SDF solutions with guidance weight
    $w=1$, which are then fed into the forward models for prediction. We select
    the four design results with the lowest $L_{2}$ relative error, shown in (c-f).
    (a) shows the comparison between the target and prediction of (c-f) using
    the forward model, with the $L_{2}$ relative error indicated in the curve legend.
    (b) shows the corresponding stress-strain curves obtained using Abaqus.}
    \label{fig:ondemand_design}
  \end{figure}

  \begin{figure}[h]
    \centering
    \includegraphics[width=\textwidth]{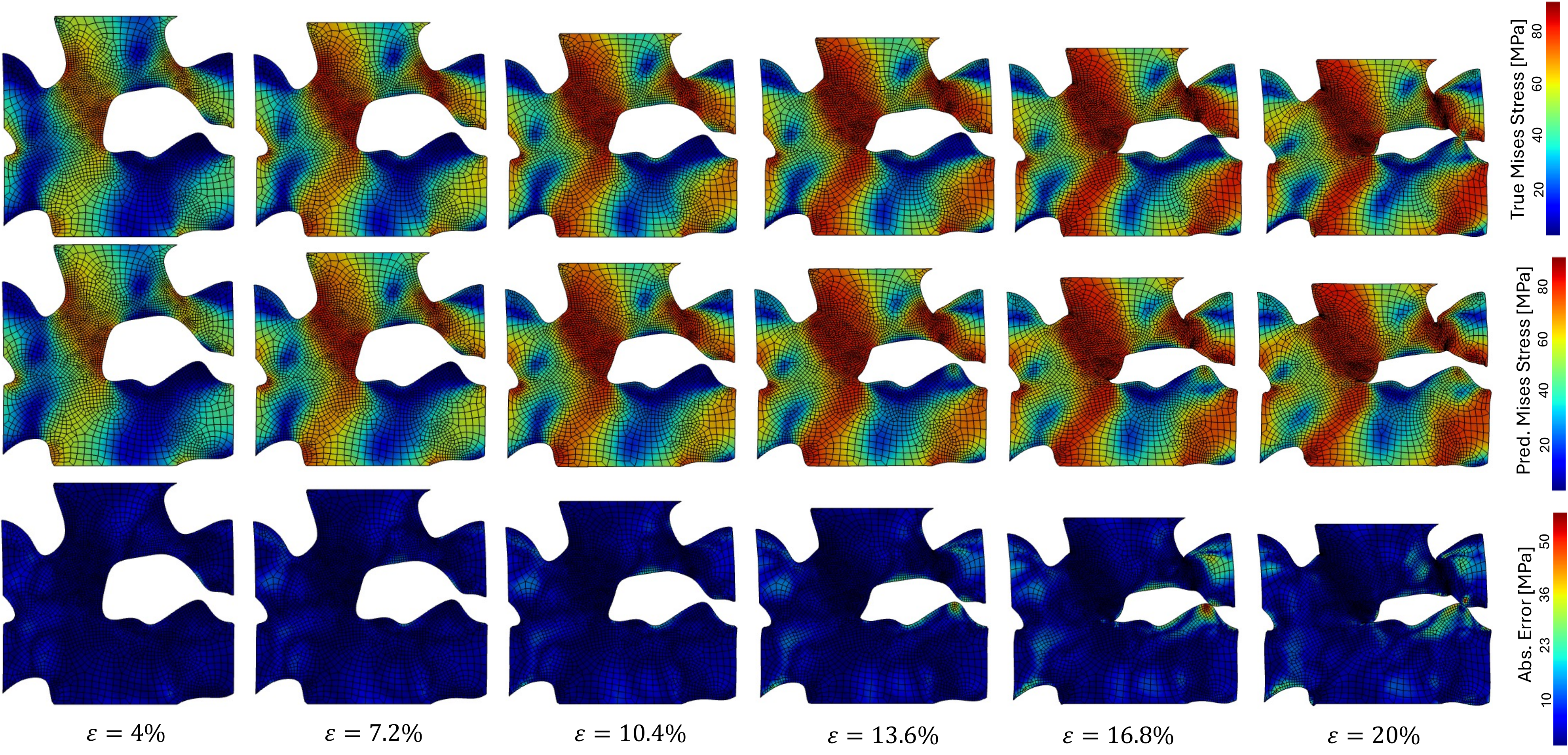}
    \caption{Comparison of the Mises stress and displacement field predictions
    with FE ground truth at different strain $\varepsilon$ steps for the
    designed structure in \cref{fig:ondemand_design}(Ae). The first row shows
    the true Mises stress on the true deformed shape, the second row shows the predicted
    Mises stress on the predicted deformed shape, and the third row shows the
    absolute error in Mises stress on the true deformed shape.}
    \label{fig:ondemand_design_SU_Ae}
  \end{figure}

  \begin{figure}[htbp]
    \centering
    \includegraphics[width=\textwidth]{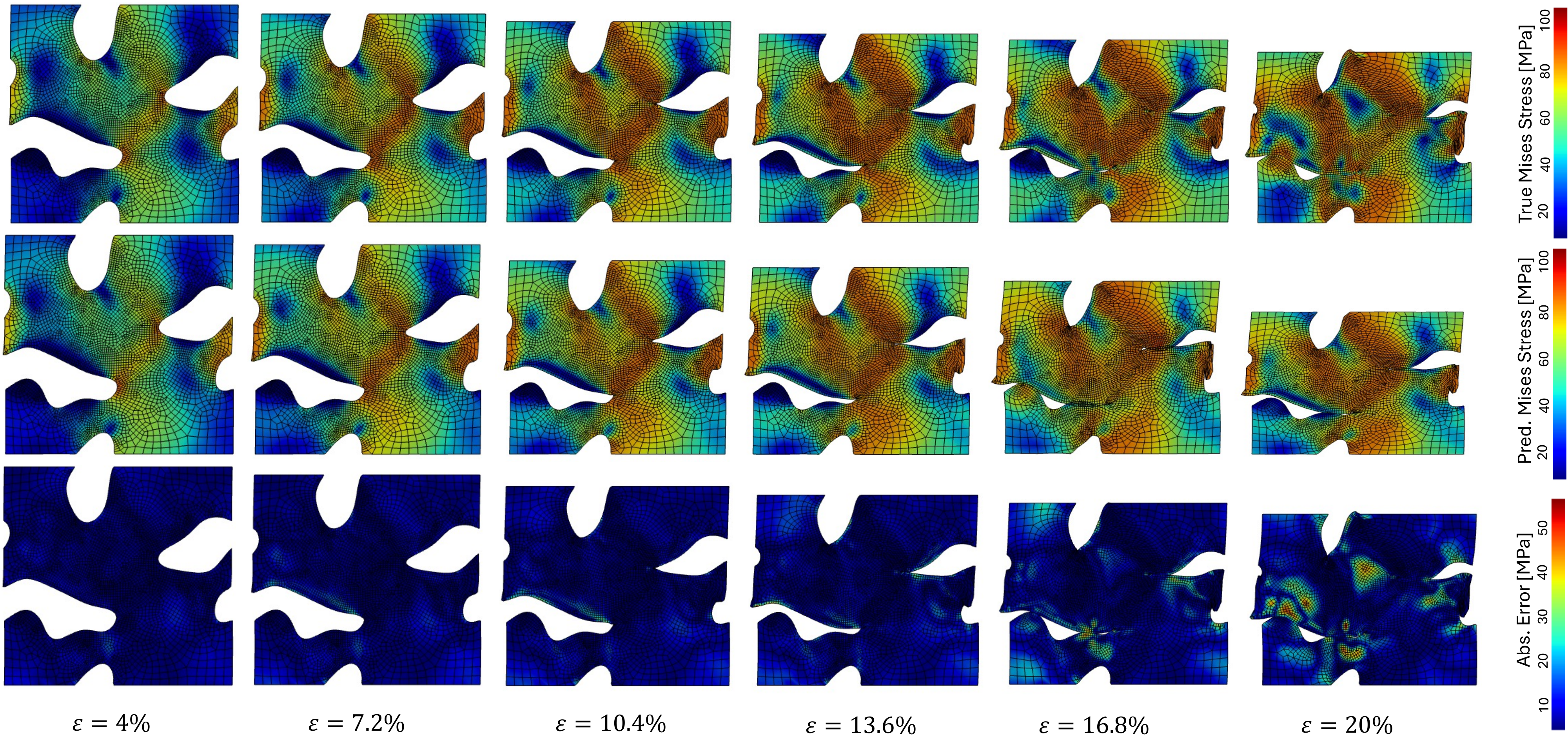}
    \caption{Comparison between the Mises stress and displacement field
    predictions with the corresponding FE ground truth at different strain
    $\varepsilon$ steps for the designed structure in \textbf{\cref{fig:ondemand_design}(Be)}.
    The first row shows the true Mises stress under the true deformed shape; the
    second row shows the predicted Mises stress under the predicted deformed shape;
    and the third row shows the absolute error of the Mises stress under the
    true deformed shape.}
    \label{fig:X_ondemand_design_SU_Be}
  \end{figure}

  While our developed diffusion model performs well on the test dataset for inverse
  retrieval, real-world applications often demand on-the-fly inverse design of
  custom-defined material properties with high fidelity. Therefore, instead of
  using stress-strain curves from FE simulations as shown in
  \cref{fig:X_test_design}, a robust inverse design model should be able to generate
  possible designs from custom and predefined stress-strain curves. Thus, we now
  use the Ramberg-Osgood equation \citep{gadamchetty2016practical} to define the
  target stress-strain curve. The Ramberg-Osgood equation is defined as
  \begin{equation}
    \label{eq:ramberg_osgood}\varepsilon=\frac{\sigma}{E}+\alpha \left( \frac{\sigma}{\sigma_{0}}
    \right )^{n},
  \end{equation}
  where $E$ is the Young's modulus, $\sigma_{0}$ is the yield stress, $\alpha$ is
  the Ramberg-Osgood coefficient, and $n$ is the hardening exponent. To mimic the
  self-contact behavior of the material, we modify the Ramberg-Osgood equation
  to include a critical strain $\varepsilon_{c}$ below which the material
  follows the Ramberg-Osgood equation and above which the material is assumed to
  behave as a linear elastic material. The modified Ramberg-Osgood equation is defined
  as
  \begin{equation}
    \label{eq:m_ramberg_osgood}\varepsilon=\left\{
    \begin{array}{ll}
      \frac{\sigma}{E}+\alpha \left( \frac{\sigma}{\sigma_{0}} \right )^{n} & \text{if}~\varepsilon \leq \varepsilon_c, \\
      \frac{\sigma-\sigma_{c}}{E'}+ \varepsilon_c                           & \text{if}~\varepsilon > \varepsilon_c,
    \end{array}
    \right.
  \end{equation}
  where $\sigma_{c}$ is the stress corresponding to $\varepsilon_{c}$ and $E'$ is
  the Young's modulus after self-contacting.

  Using the Ramberg-Osgood equations, we generate two target stress-strain
  curves with different material properties, as shown in \cref{fig:ondemand_design}(A)
  and (B). Both targets use the same hardening exponent $n=10$ and coefficient $\alpha
  = 0.002$, but differ in Young's modulus and reference yield stress. Case (A) is
  obtained using \cref{eq:ramberg_osgood} with a Young's modulus $E=800$ MPa and
  yield stress $\sigma_{0}=30$ MPa, while Case (B) is derived from
  \cref{eq:m_ramberg_osgood} with $E=1000$ MPa, $\sigma_{0}=40$ MPa, $E'=500$
  MPa, and $\varepsilon_{c}=0.156$. These target curves are fed into the inverse
  diffusion model to generate 500 geometry solutions, which are then evaluated using
  the forward model to predict the stress-strain curves. The four best designs
  based on the $L_{2}$ relative error are shown in \cref{fig:ondemand_design} (c-f).
  \cref{fig:ondemand_design}(a) and (b) compare the target and designed stress-strain
  curves for the four geometries using the forward model and FE simulations, respectively,
  with the $L_{2}$ relative error indicated in the curve legend. As shown in \cref{fig:ondemand_design}(a),
  the designed geometries closely match the custom-defined stress-strain curves.
  The FE simulations of the designed geometries, shown in \cref{fig:ondemand_design}(b),
  confirm the accuracy of the designs. Additionally, we predict the solution fields
  for the designed geometries \cref{fig:ondemand_design}(Ae) and \cref{fig:ondemand_design}(Be) at different
  strain steps, with results shown in \cref{fig:ondemand_design_SU_Ae} and
  \cref{fig:X_ondemand_design_SU_Be}, respectively. These results demonstrate
  that the designed geometries achieve the target stress-strain curves and
  accurately predict the solution fields.

  \section{Methods}
  \label{sec:methods} This section provides details on the generation of random
  unit cell geometries, FE simulations, and the implementation of deep learning
  models. Additional explanations can be found in the Supplementary Information.

  \subsection{Unit cell geometries}
  To generate periodic unit cells, Bastek et al. \citep{bastek2023inverse}
  generate one-quarter of the unit cell represented by binary pixels and then
  mirror the other three quarters, which is a common practice in the literature but
  results in bi-axis symmetric and not fully arbitrary unit cells. Additionally,
  such binary pixel-based geometries have unsmooth boundaries (\cref{fig:S_geo_rep}(a)),
  which cannot be directly applied for simulation or 3D printing without a boundary
  smoothing process. Instead, we generate a random metamaterial unit cell by sampling
  a periodic 2D Gaussian random field $U\left(x\right)$ on a square domain $[0,1~
  \text{mm}]^{2}$ with a resolution of $64\times 64$, incorporating the periodic
  Fourier method \citep{Muller2022}:
  \begin{equation}
    U\left(x\right)= \sum_{i=1}^{N}\sqrt{2S(k_{i})\Delta k}\left( Z_{1,i}\cdot\cos
    \left( \textbf{k}_{i}\cdot \textbf{x}\right)+ Z_{2,i}\cdot\sin\left( \textbf{
     k}_{i}\cdot \textbf{x}\right) \right),
  \end{equation}
  where $S$ is the spectrum of the Gaussian covariance model, $Z_{1,i}, Z_{2,i}\sim
  N(0,1)$ are mutually independent and drawn from a standard normal distribution,
  and $\textbf{k}_{i}$ is the equidistant Fourier grid which ensures the
  periodic field. This algorithm is implemented in GSTools \citep{mueller_schueler_2018},
  a Python-based open-source package. The generated Gaussian random field (GRF)
  is then thresholded at a random value to extract the contour using a marching
  algorithm, with the extracted contour defining the unit cell boundary of the metamaterial.
  The GRF values less than the threshold are set as void and the extracted
  closed contours form the internal boundaries of holes. Unclosed contours intersect
  with the boundary of the square domain and the segment between the two intersecting
  points on the same edge of the square domain are connected as additional unclosed
  boundary contours if the GRF values between the two points are higher than the
  threshold. Connecting the unclosed contours forms the external boundary of the
  unit cell. We exclude the generated unit cells containing ``islands" and also
  ensure that opposite boundaries of the domain have enough distance to connect
  with neighboring unit cells by excluding samples with small edge lengths (less
  than 8\% of domain size). Small internal holes are removed to ensure the
  geometry is printable. To ensure high mesh quality in Abaqus simulations,
  adjacent points on a contour are merged if the distance is too close. Using this
  geometry generation method, we generate about 10,000 valid arbitrary random
  periodic unit cell geometries. We evaluate the shortest distance of $120 \times
  120$ uniform grid points in $[-0.1~\text{mm},1.1~\text{mm}]^{2}$ to the boundary
  contours for the SDF representation of the geometry.

  \subsection{FE simulation}
  \label{sec:fesim} We use the commercial finite elements code Abaqus 2024 to simulate
  the material response of the generated unit cell geometries under compression.
  The nodes along the bottom edge are fixed along the y-direction ($u_{y}=0$)
  and the nodes along the top edge are subjected to displacement control with a
  maximum compressive strain of 0.2. Periodic boundary conditions are applied along
  the left and right sides of the unit cell. The horizontal motion $u_{x}$ of
  the nearest node to the center of the unit cell is set at zero to remove rigid
  body motion. Similar to ref.\citep{bastek2023inverse}, we apply self-contact with
  a friction coefficient of 0.4 to prevent the unit cell from collapsing during compression.

  An elasto-plastic material model with large deformation is used for the simulations,
  with material properties detailed in Supplementary \cref{s_sec:mat_prop}. An implicit
  dynamic solver with virtual mass density $\rho=10^{-8}$ is used to mimic quasi-static
  compression while allowing better convergence for large deformation and elasto-plastic
  models \citep{bastek2023inverse}. The compression strain is applied progressively
  in 51 steps. Mixed 1st-order quadrilateral elements and 2nd-order triangular
  elements with full integration and plane-strain assumption are used for the
  simulations. The global mesh size is set to 0.04 mm. Note that the mesh size
  around the boundary contour is approximately 1/64 mm, which is close to the segment
  length of boundary contours, as the geometry contour is extracted from a GRF
  with a resolution of $64\times64$. We performed a mesh sensitivity analysis and
  found that a mesh size of 0.04 mm is sufficient to capture the material response
  of the unit cell, as presented in Supplementary \cref{s_sec:mesh_sen}. We first
  use Abaqus CAE with Python API to generate the input files for the Abaqus solver
  for all geometries and then use the Abaqus solver to perform the simulations based
  on these input files. All computational tasks of data generation are performed
  on the DELTA machine at the National Center for Supercomputing Applications (NCSA)
  at the University of Illinois Urbana-Champaign.

  \subsection{Deep learning models}
  In this section, we provide a high-level summary of the model architecture.
  Further implementation details can be found in the supplementary information
  and our GitHub repository. We developed two forward neural operator transformers
  (NOT) for predicting the stress-strain curve and the solution fields for arbitrary
  geometries represented by SDF. The SDF is represented by a 2D image with a
  shape of $120 \times 120$. A 2D U-Net architecture is used to encode the SDF.
  The U-Net maintains the image size of $120 \times 120$ as output, followed by three
  down-sampling layers to reduce the image size. After reshaping, the output of the
  geometry encoder is a matrix with a shape of $15^{2}\times d_{e}$, where
  $d_{e}$ is the embedding dimension.

  The other inputs to NOT are the query points, which are the strain steps with a
  shape of $N \times 1$ or the node coordinates with a shape of $N \times 2$ for
  the two NOT models, respectively, in which $N$ is the number of query points.
  For the strain steps, $N=51$. Since the geometry is arbitrary, the number of mesh
  nodes varies for each geometry. To allow for batch processing, the mesh nodes are
  padded to the maximum number of nodes in the batch. The NeRF positional
  encoding followed by an MLP is applied to the query points, whose output has a
  shape of $N \times d_{e}$, which is then fused with the output of the geometry
  encoder using the attention mechanism. Residual connections and layer normalization
  are applied in each attention block to stabilize the training. The output of a
  few attention blocks is decoded using an MLP to the solution on the query
  points, e.g., the effective stress with a shape of $N \times 1$ and the 26 frames
  of solution fields (Mises stress and two displacements) with a shape of $N \times
  (3 \times 26)$ for the two NOT models, respectively.

  For the inverse diffusion model, we generate the SDF from pure Gaussian noise on
  uniform, fixed grids, which is a common image-to-image translation task. The
  residual U-Net is a prevalent choice for such tasks and we use a similar
  architecture to the original paper of the diffusion model, which applies the attention
  mechanism at the bottom two levels of the U-Net. To incorporate the condition
  of the target stress-strain curve with a shape of $51 \times 1$ into the noise
  estimator, we use an MLP to encode the target curve and then expand and repeat
  the output to the same shape as the noise image, so that it can be
  concatenated with the image at each level of the U-Net. The output of the diffusion
  model is the SDF with a shape of $120 \times 120$, which is then fed into the
  marching algorithm to extract the geometry.

  \subsection{Training protocol}
  Before training, we normalize the SDF, stress, and displacement to zero mean and
  unit variance. The dataset is split into 80\% for training and 20\% for
  testing. The mean square error is used as the loss function. We implement this
  work using the PyTorch framework, and training is performed on a single NVIDIA
  H100 GPU with 90GB memory on the DeltaAI machine available at NCSA. We use the
  Adam optimizer \citep{kingma2014adam} with an initial learning rate of $0.001$,
  leveraging its adaptive moment estimation for stable and efficient convergence.
  To further enhance training stability and adapt the learning rate dynamically,
  we incorporate the \emph{ReduceLROnPlateau} scheduler, which reduces the
  learning rate when the validation loss stagnates. For padding query points, a
  mask mechanism is applied to exclude them from the loss calculation. The loss value
  evolution during training is shown in Supplementary Information (\cref{fig:S_notss_perform} and \cref{fig:S_notsu_perform}.)
  Detailed training information is provided in the Supplementary Information. Specifically,
  the computational efficiency is discussed in Supplementary
  \cref{s_sec:efficiency}.

  \subsection{Inference protocol}
  All inference tasks are performed on a single NVIDIA A100 GPU on the DELTA
  machine at NCSA.

  For the inverse diffusion model, we generate multiple geometries for a single target
  stress-strain curve. We apply a filter to remove the samples with ``islands" and
  ensure that the geometry is periodic. The filtered geometry is then fed into
  the forward model to predict the stress-strain curve. The predicted stress-strain
  curve is compared with the target curve using the $L_{2}$ relative error.

  \subsection{Metrics}
  We use the $L_{2}$ relative error as the metric to evaluate the performance of
  the models. For the stress-strain curve, the $L_{2}$ relative error is defined
  as
  \begin{equation}
    L_{2}=\frac{\left\| \sigma_{eff}^{true}-\sigma_{eff}^{pred}\right\|_{2}}{\left\|
    \sigma_{eff}^{true}\right\|_{2}},
  \end{equation}
  where $\left\| \cdot \right\|_{2}$ is the $L_{2}$ norm. For the solution
  fields, the $L_{2}$ relative error is defined as
  \begin{equation}
    L_{2}=\frac{1}{3}\left(\frac{\left\| \sigma_{mises}^{true}-\sigma_{mises}^{pred}\right\|_{2}}{\left\|
    \sigma_{mises}^{true}\right\|_{2}}+\frac{\left\| u_{x}^{true}- u_{x}^{pred}\right\|_{2}}{\left\|
    u_{x}^{true}\right\|_{2}}+\frac{\left\| u_{y}^{true}-u_{y}^{pred}\right\|_{2}}{\left\|
    u_{y}^{true}\right\|_{2}}\right),
  \end{equation}
  where the $L_{2}$ relative error is calculated over all the strain steps.

  \section{Conclusions}
  The integration of additive manufacturing and advanced computational techniques
  has opened new avenues for the design and synthesis of metamaterials with tailored
  mechanical properties with applications in various engineering disciplines including
  aerospace, automotive, and biomedical. One of the primary challenges addressed
  in this work is the complexity of inverse design in achieving desired
  macroscopic mechanical properties, especially in nonlinear scenarios.
  Traditional methods, such as topology optimization, although effective, are often
  computationally intensive and prone to local minima due to their reliance on iterative
  gradient-based optimization schemes. Recently, generative deep learning models
  have shown potential in inverse design by offering a more efficient approach that
  eliminates the need for multiple iterations, thus providing a pathway for
  exploring the vast design space of metamaterials. However, existing methods
  typically rely on binary pixel-based geometries, which have jagged boundaries and
  require additional boundary smoothing, and are thus not directly applicable to
  FE simulations and 3D printing. In contrast, our proposed method designs
  architected metamaterials using a classifier-free guided diffusion model, representing
  geometries through signed distance functions (SDFs). This approach enables the
  generation of multiple SDF solutions for a given target stress-strain curve. The
  transition from binary pixel-based representations to an SDF framework marks a
  significant advancement, as SDFs provide smooth boundary representations that
  eliminate jagged edges. This refinement reduces mesh distortions in finite element
  simulations and simplifies the 3D printing process, ultimately improving
  simulation accuracy, convergence, and manufacturability, and ensuring the robustness
  and reliability of the designed structures.

  Real-world applications often require on-the-fly design adaptations to achieve
  custom material properties. Our results demonstrate the ability of the
  proposed framework to generate diverse and feasible designs that closely align
  with custom target stress-strain curves. The stochastic nature of generative diffusion
  models allows for multiple solutions to a single target, but efficiently
  evaluating the performance and mechanical behavior of these designs remains a
  challenge. To address this issue, the forward Neural Operator Transformer (NOT)
  developed in this work enhances the inverse design framework by enabling
  accurate prediction of the macroscopic stress-strain curves and the local
  stress and displacement fields for arbitrary designed geometries. This capability
  provides deeper insight into the mechanical properties of the designed structures.
  Additionally, NOT's adaptability to varying mesh configurations strengthens its
  applicability, making it a robust prediction tool capable of accommodating the
  irregularities inherent in real-world applications.

  The proposed closed-loop framework, which integrates inverse design and forward
  prediction models, offers a versatile and efficient approach for designing and
  synthesizing metamaterials with tailored mechanical properties. Beyond predicting
  homogenized stress-strain curves, the developed forward neural operator
  transformer (NOT) models can efficiently compute solution fields for arbitrary
  geometries. This framework can be naturally extended by conditioning the model
  not only on target stress-strain curves but also on additional material properties,
  such as maximum stress, which is crucial for assessing material damage.
  Expanding the framework primarily requires additional computational efforts, including
  data generation and model training, while the developed models remain directly
  applicable.

  \section{Data availability}
  The dataset is available on Zenodo \citep{liu2025dataset} and the trained models
  are available at the Github repository \url{https://github.com/QibangLiu/SDFGeoDesign}

  \section{Code availability}
  The codes for training and inference are available at the Github repository
  \url{https://github.com/QibangLiu/SDFGeoDesign}

  \section{Acknowledgements}
  The authors would like to thank the National Center for Supercomputing Applications
  (NCSA) at the University of Illinois Urbana-Champaign, and particularly its
  Research Computing Directorate, Industry Program, and Center for Artificial
  Intelligence Innovation (CAII) for support and hardware resources. This research used both the DeltaAI advanced computing and data resource, which is supported by the National Science Foundation (award OAC 2320345) and the State of Illinois, and the Delta advanced computing and data resource which is supported by the National Science Foundation (award OAC 2005572) and the State of Illinois. Delta and DeltaAI are joint efforts of the University of Illinois Urbana-Champaign and its National Center for Supercomputing Applications.

  \section{Author contributions statement}
  Q. Liu conceived the concept, generated the dataset, developed the methodology,
  implemented the models, performed the analysis, and drafted, reviewed, and edited
  the manuscript. S. Koric contributed to the results discussion and to the configuration
  of Abaqus on HPC, and reviewed and edited the manuscript. D. Abueidda and H. Meidani
  contributed to the result discussion and reviewed and edited the manuscript. P.
  Geubelle proposed the concept, supervised the project, contributed to the result
  discussion, and reviewed and edited the manuscript.

  \section{Competing interests statement}
  The authors declare no competing financial or non-financial interests.

  \bibliographystyle{elsarticle-num-names}
  \bibliography{reference}%\FloatBarrier
  \newpage
  \captionsetup[figure]{labelfont={bf},name={Extended Data Fig.},labelsep=colon}
  % \captionsetup [table]{labelfont={bf},name={Extended Data Table},labelsep=colon}
  \setcounter{figure}{0}% \input{ED}

  % ondemand_design
  %%%%%%%%%%%%%%  %%%%%%%%%%%%%%  %%%%%%%%%%%%%%  %%%%%%%%%%%%%%
  %%%%%%%%%%%%%%  Supplementary information   %%%%%%%%%%%%%%
  %%%%%%%%%%%%%%  %%%%%%%%%%%%%%  %%%%%%%%%%%%%%  %%%%%%%%%%%%%%
  \clearpage
  \captionsetup[figure]{labelfont={bf},name={Fig.},labelsep=colon}
  \setcounter{equation}{0}
  \setcounter{section}{-1}
  \setcounter{figure}{0}
  \setcounter{table}{0}
  \setcounter{page}{1}% set equations to begin with S
  \renewcommand{\theequation}{S\arabic{equation}}% set page to begin with S
  \renewcommand{\thepage}{S\arabic{page}}% set sections to begin with Supplementary Section
  \renewcommand{\thesection}{S\arabic{section}}% set figures to begin with Supplementary figure
  \renewcommand{\thefigure}{S\arabic{figure}}% \input{SI}
  \renewcommand{\thetable}{S\arabic{table}}
  \begin{center}
    \Large Supplementary information for ``Toward Signed Distance Function based
    Metamaterial Design: Neural Operator Transformer for Forward Prediction and
    Diffusion Model for Inverse Design" \break

    \normalsize Qibang Liu$^{1,6,*}$, Seid Koric$^{1,4}$, Diab Abueidda$^{1,7}$,
    Hadi Meidani$^{5}$, Philippe Geubelle$^{2,3,**}$ \break

    ${}^{1}$\NCSA \\ ${}^{2}$\BI\\ ${}^{3}$\AE, \\ ${}^{4}$\MSE\\ ${}^{5}$\CEE\\
    ${}^{6}$\KSU\\ ${}^{7}$\NYU\\ ${}^{*}$Corresponding author: Q. Liu, qibang@illinois.edu\\
    ${}^{**}$Corresponding author: P. Geubelle, geubelle@illinois.edu
  \end{center}
  \break

  \section{Overview}
  We here provide further details regarding the data generation procedure, model
  implemtation, and futher results. More details can be found in the Github
  repository \url{https://github.com/QibangLiu/SDFGeoDesign}

  \section{Binary pixel-based vs. SDF-based geometry representation}
\begin{figure}[ht]
    \centering
    \includegraphics[width=\textwidth]{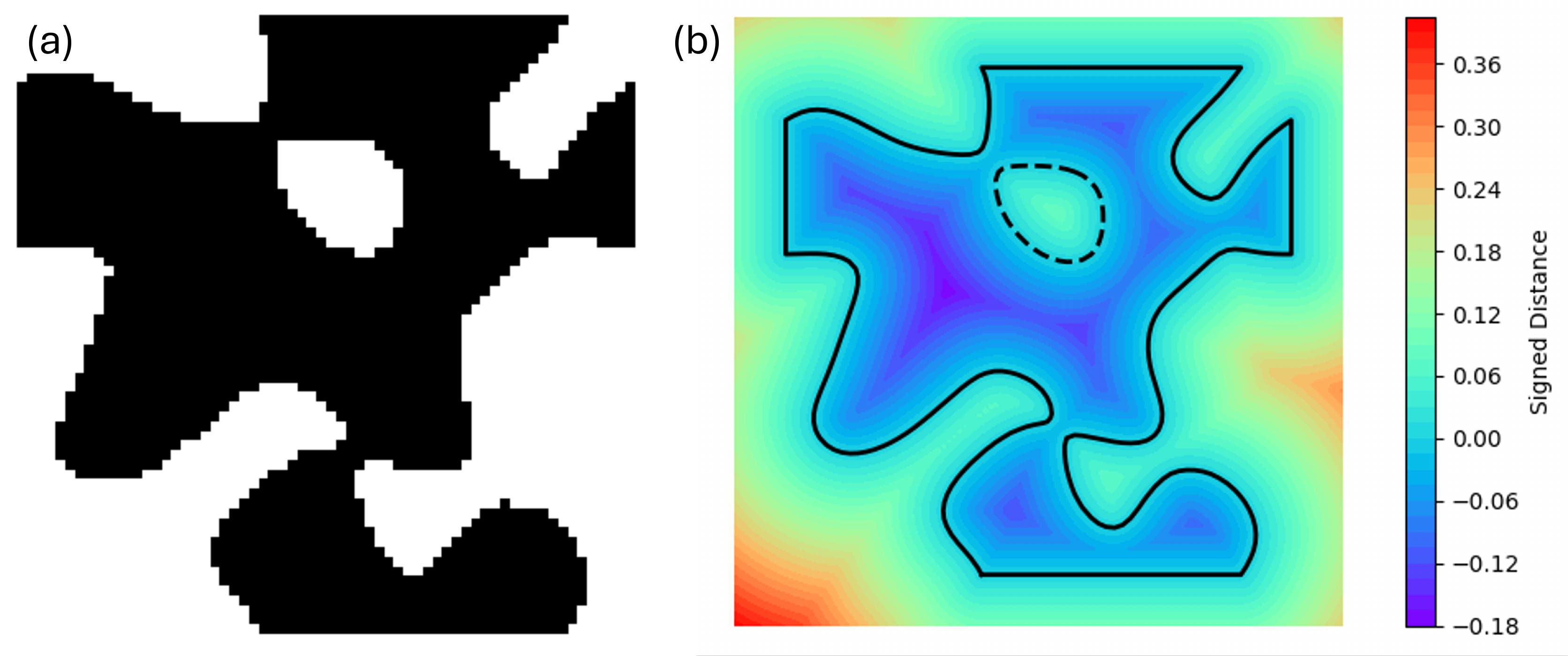}
    \caption{Geometry represented by pixel density (a) and by SDF (b).}
    \label{fig:S_geo_rep}
  \end{figure}

  The binary pixel-based geometry representation is commonly used for metamaterial
  design, where the geometry is represented by a binary image with pixel values
  of 0 or 1. As shown in \cref{fig:S_geo_rep}(a), this representation is simple
  and intuitive but has limitations, such as jagged boundaries and the need for additional
  boundary smoothing. In contrast, the signed distance function (SDF) representation,
  shown in \cref{fig:S_geo_rep}(b), provides a smooth boundary representation
  that eliminates jagged edges and simplifies the finite element simulation and
  3D printing process. The SDF representation is defined as the shortest distance
  from a point in space to the boundary of the geometry, with negative values
  inside the geometry, positive values outside the geometry, and zero values on
  the boundary. This continuous and differentiable representation is suitable for
  neural network-based design and simulation tasks.

  \section{Data generation}

  \subsection{Periodic unit cell}
  We generate a random periodic unit cell by sampling a periodic 2D Gaussian
  random field $U\left(x\right)$ on a square domain $[0,1~\text{mm}]^{2}$ with a
  resolution of $64\times 64$, using the periodic Fourier method \citep{s_muller2022}:
  \begin{equation}
    U\left(x\right)= \sum_{i=1}^{N}\sqrt{2S(k_{i})\Delta k}\left( Z_{1,i}\cdot\cos
    \left(\left\langle k_{i},x\right\rangle \right)+ Z_{2,i}\cdot\sin\left(\left\langle
    k_{i},x\right\rangle \right) \right),
  \end{equation}
  where $S$ is the spectrum of the Gaussian covariance model,
  $Z_{1,i}, Z_{2,i}\sim N(0,1)$ are mutually independent and drawn from a
  standard normal distribution, and $k_{i}$ is the equidistant Fourier grid. We
  use the open-source Python package GSTools \citep{s_mueller_schueler_2018} to generate
  such 2D Gaussian random fields. For the Gaussian covariance model, we set the variance
  to 10 and the length scale to 0.15. For the Fourier space, we use $32^{2}$ modes.
  The GRF is evaluated as follows:

  \begin{lstlisting}[language=Python]
    x = np.linspace(0, 1, 64)
    y = np.linspace(0, 1, 64)
    model = gstools.Gaussian(dim=2, var=10, len_scale=0.15)
    srf = gstools.SRF(
      model,
      generator="Fourier",
      period=(1,1),  # periodic in x and y
      mode_no=32,
      seed=None)
    grf=srf((x, y), mesh_type="structured")
  \end{lstlisting}
  For the implementation of the 2D marching algorithm, we use the open-source Python
  package {\fontfamily{qcr}\selectfont scikit-image}.

  For each generated unit cell, we store the coordinates of the vertices and the
  contour connectivities (outer boundary and holes) in a json file, which can be
  read by Abaqus using the Python API for building geometry and meshing. To
  evaluate the SDF for each unit cell, we first build a polygon object using {\fontfamily{qcr}\selectfont shapely.geometry}
  from the open-source Python package {\fontfamily{qcr}\selectfont shapely},
  then calculate the shortest distance of $120 \times 120$ uniform grid points
  in $[-0.1~\text{mm}, 1.1~\text{mm}]^{2}$ to the polygon.

  \subsection{Material properties}
  \label{s_sec:mat_prop} 
\begin{figure}[!h]
    \centering
    \includegraphics[width=4in]{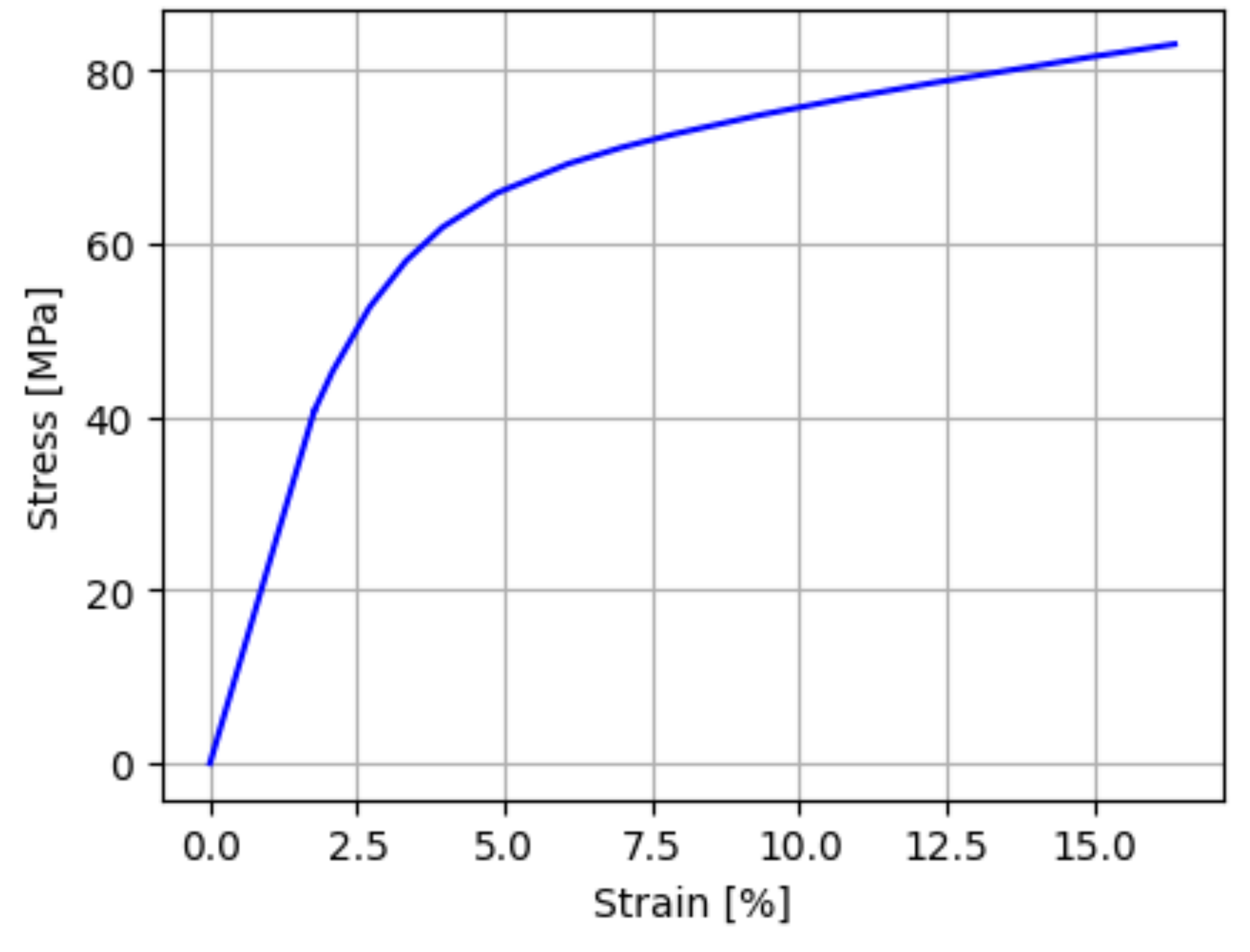}
    \caption{Stress-strain curve for homopolymer resin.}
    \label{fig:S_mat}
  \end{figure}

The material adopted in this work is acetal homopolymer
  resin \citep{jin2020guided} and an elastic-plastic model is used to describe its
  constitutive behavior. The Young's modulus of the resin is 2.3 GPa and the
  Poisson's ratio is 0.35. The plastic behavior is described by a piecewise linear
  hardening model presented in \cref{tab:mat_prop} and the corresponding stress-strain
  curve is shown in \cref{fig:S_mat}.
  \begin{table}[hb]
    \centering
    \caption{Material properties of the homopolymer resin.}
    \begin{tabular}{c|ccccccc}
      \hline
      Plastic strain [\%] & 0.0    & 0.1133 & 0.4183 & 0.80645 & 1.2557 & 2.0035 & 3.0689 \\
                          & 3.8873 & 4.7114 & 6.083  & 7.4477  & 8.799  & 11.457 & 12.76  \\
      \hline
      Stress [MPa]        & 40.62  & 45.24  & 52.62  & 58.00   & 61.87  & 65.81  & 69.19  \\
                          & 71.06  & 72.61  & 74.82  & 76.74   & 78.46  & 81.58  & 83.00  \\
      \hline
    \end{tabular}
    \label{tab:mat_prop}
  \end{table}

  \subsection{Mesh sensitivity analysis}
  \label{s_sec:mesh_sen}

\begin{figure}[!h]
    \centering
    \includegraphics[width=6in]{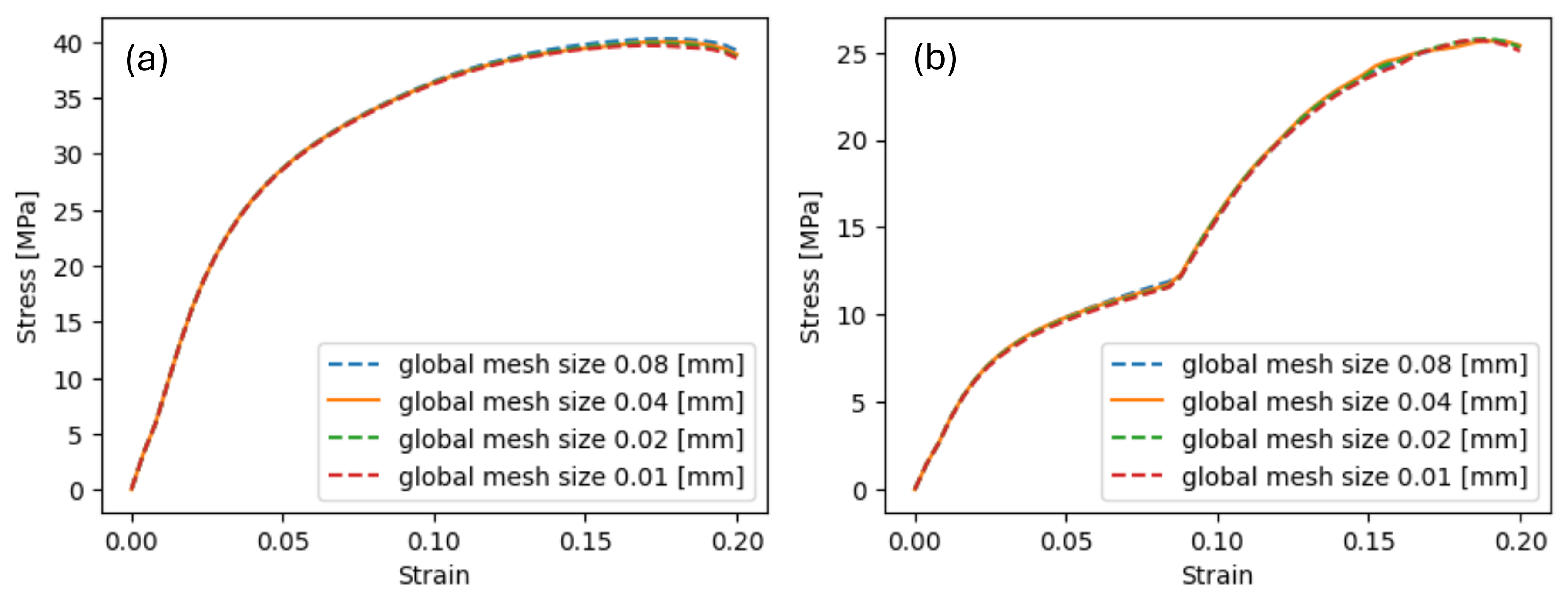}
    \caption{Mesh sensitivity analysis for two unit cells under compression: (a)
    without self-contact and (b) with self-contact.}
    \label{fig:S_mesh_sen}
  \end{figure}

  We conducted a mesh sensitivity analysis to determine the optimal mesh size for
  the finite element simulations. It is important to note that the domain
  boundary consists of small segments, each approximately 1/64 mm in length.
  This segment length corresponds to the resolution of the Gaussian random field
  (GRF) used to generate the geometry. Consequently, the mesh size around the
  contour does not exceed the segment length. Our analysis indicates that a global
  mesh size of 0.04 mm is sufficient to capture the compression response of the unit
  cells. The stress-strain curves obtained using a mesh size of 0.04 mm are
  compared with the curves obtained using mesh sizes of 0.01 mm, 0.02 mm, and
  0.08 mm as shown in \cref{fig:S_mesh_sen}. The results indicate that the stress-strain
  curves obtained using the four mesh sizes are almost identical, confirming
  that a mesh size of 0.04 mm is adequate to capture the material response of
  the unit cell.

  \subsection{Data augementation}
  Based on the random, arbitrary periodic unit cell generation and the
  corresponding FE simulation, we generate 10K samples. Training the forward
  neural networks exhibits severe overfitting due to the limited number of
  training samples and the complexity of the data. To alleviate this issue, we
  apply data augmentation to the dataset by randomly shifting the unit cell in
  the $x$-direction multiple times. The new unit cell $[x_{0}, x_{0}+1 ]\times [0
  , 1] ~\text{mm}^{2}$ is created by shifting the original unit cell horizontally
  by $x_{0}$, where $x_{0}$ is randomly selected from $[0.1, 0 .9]$ mm. The solution
  stress and displacement fields of the new unit cell are the same as the original
  one but with the node coordinates shifted by $x_{0}$:
  \begin{equation}
    x_{\text{new}}=
    \begin{cases}
      x_{\text{original}}+ 1 - x_{0} & \text{if }x_{\text{original}}< x_{0},    \\
      x_{\text{original}}- x_{0}     & \text{if }x_{\text{original}}\geq x_{0}.
    \end{cases}
  \end{equation}
  The stress-strain curve of the new unit cell remains the same as the original
  one due to the periodic boundary condition. For each unit cell, we apply 8
  random shifts. After filtering out unit cells with ``islands", we obtain a total
  of 73,879 training samples.

  \clearpage

  \section{Forward models}
  \label{s_sec:forward}

  \subsection{Model architectures}
  The forward neural operator transformer (NOT) model is designed to predict the
  stress-strain curve and the solution fields (Mises stress and two
  displacements). We summarize the most relevant hyperparameters of the two
  forward NOT models in this section. The NOT model consists of two main components:
  a geometry encoder and a solution decoder. The geometry encoder encodes the geometry
  represented by the signed distance function (SDF) into a latent space, which
  is fed into attention blocks as the Key and Value. The geometry encoder is summarized
  in \cref{tab:geoencoder}.

  \begin{table}[!h]
    \centering
    \caption{Geometry encoder architecture of the forward NOT model.}
    \begin{tabular}{|c|l|l|}
      \hline
      \textbf{Layer \#} & \textbf{Layer type (Description)}                        & \textbf{Output shape} \\
      \hline
      1                 & Input                                                    & (B, 1, 120, 120)      \\
      \hline
      2                 & Conv2D                                                   & (B, 8, 120, 120)      \\
      3                 & SiLU+Conv2d+GroupNorm+SiLU+Conv2D (Residual Conv Block)  & (B, 8, 120, 120)      \\
      4                 & Conv2d (DownSample)                                      & (B, 8, 60, 60)        \\
      5                 & SiLU+Conv2d+GroupNorm+SiLU+Conv2D (Residual Conv Block)  & (B, 16, 60, 60)       \\
      6                 & Conv2d (DownSample)                                      & (B, 16, 30, 30)       \\
      7                 & SiLU+Conv2d+GroupNorm+SiLU+Conv2D (Residual Conv Block)  & (B, 32, 30, 30)       \\
      8                 & Conv2d (DownSample)                                      & (B, 32, 15, 15)       \\
      9                 & SiLU+Conv2d+GroupNorm+SiLU+Conv2D (Residual Conv Block)  & (B, 64, 15, 15)       \\
      \hline
      10                & SiLU+Conv2d+GroupNorm+SiLU+Conv2D (Residual Conv Block)  & (B, 64, 15, 15)       \\
      11                & Self-Attention Block                                     & (B, 64, 15, 15)       \\
      12                & SiLU+Conv2d+GroupNorm+SiLU+Conv2D (Residual Conv Block)  & (B, 64, 15, 15)       \\
      \hline
      13                & Concatenate                                              & (B, 128, 15, 15)      \\
      14                & SiLU+ Conv2d+GroupNorm+SiLU+Conv2D (Residual Conv Block) & (B, 64, 15, 15)       \\
      15                & Conv2D (UpSample)                                        & (B,64,30,30)          \\
      16                & Concatenate                                              & (B, 96, 30, 30)       \\
      17                & SiLU+ Conv2d+GroupNorm+SiLU+Conv2D (Residual Conv Block) & (B, 32, 30, 30)       \\
      18                & Conv2D (UpSample)                                        & (B, 32, 60, 60)       \\
      19                & Concatenate                                              & (B, 48, 60, 60)       \\
      20                & SiLU+Conv2d+GroupNorm+SiLU+Conv2D (Residual Conv Block)  & (B, 16, 60, 60)       \\
      21                & Conv2D (UpSample)                                        & (B, 16, 120, 120)     \\
      22                & Concatenate                                              & (B, 24, 120, 120)     \\
      23                & SiLU+Conv2d+GroupNorm+SiLU+Conv2D (Residual Conv Block)  & (B, 8, 120, 120)      \\
      \hline
      24                & Conv2D+SiLU                                              & (B, 64, 60, 60)       \\
      25                & Conv2D+SiLU                                              & (B, 64, 30, 30)       \\
      26                & Conv2D+SiLU                                              & (B, 64, 15, 15)       \\
      27                & Reshape + Permute (Output)                               & (B,225,64)            \\
      \hline
    \end{tabular}

    \label{tab:geoencoder}
  \end{table}

  \begin{table}[!h]
    \centering
    \caption{Solution decoder architecture of the forward NOT model. $N$ represents
    the number of query points. For the stress-strain curve prediction, $N=51$,
    while for solution fields prediction, $N$ represents the maximum number of sampled
    nodes of in the batch. $C_{in}$ represents the input channels and is equal
    to 1 for stress-strain curve prediction and 2 for solution fields prediction.
    $C_{out}$ corresponds to the output channels and is equal to 1 for stress-strain
    curve prediction, and $3 \times 26$ for solution field prediction (Mises stress
    and displacements at 26 strain steps).}
    \begin{tabular}{|c|l|l|}
      \hline
      \textbf{Layer \#} & \textbf{Layer type (Description)}                                                                  & \textbf{Output shape} \\
      \hline
      1                 & Input                                                                                              & (B, N, $C_{in}$)      \\
      \hline
      2                 & Nerf Positional Encoding+Linear+ReLU                                                               & (B, N, $C_{in}*31$)   \\
      3                 & MLP with ReLU (channels: $C_{in}*31\rightarrow 192 \rightarrow 256 \rightarrow 128 \rightarrow64$) & (B, N, 64)            \\
      4                 & Cross Attension (K,V:(B,225,64), Q(B,N,64) )                                                       & (B, N, 64)            \\
      5                 & Cross Attension (K,V:(B,225,64), Q(B,N,64) )                                                       & (B, N, 64)            \\
      6                 & MLP with ReLU (channels: $64\rightarrow128\rightarrow256\rightarrow256$)                           & (B, N, 256)           \\
      \hline
      7                 & Linear (Output)                                                                                    & (B, N, $C_{out}$)     \\
      \hline
    \end{tabular}
    \label{tab:sol_decoder}
  \end{table}
  \newpage

  \subsection{Loss function}
  The mean square error (MSE) is used as the loss function for training the forward
  NOT models. The loss function is defined as
  \begin{equation}
    \text{MSE}=\frac{1}{N}\sum_{i=1}^{N}\left( y_{i}-\hat{y}_{i}\right)^{2}.
  \end{equation}
  When training the forward NOT model for predicting the solution fields, as the
  number of query points is padded to the maximum number of nodes in the batch,
  we apply a mask mechanism to exclude the padding points from the loss
  calculation:
  \begin{equation}
    \text{MSE}=\frac{1}{\sum_{i=1}^{N}m_{i}}\sum_{i=1}^{N}m_{i}\left( y_{i}-\hat{y}
    _{i}\right)^{2},
  \end{equation}
  where $m_{i}$ is the mask value, which is 1 if it is not a padded point, and 0
  if it is. The training hyperparameters for the forward NOT models are
  summarized in \cref{tab:not_train_hyper}.

  \begin{table}[!h]
    \centering
    \caption{Training hyperparameters of the forward NOT models.}
    \begin{tabular}{c|c}
      \hline
      \textbf{Hyperparameter}          & \textbf{Value}    \\
      \hline
      Batch size (full field)          & 64                \\
      Batch size (stress-strain curve) & 128               \\
      Initial learning rate            & 1e-3              \\
      Optimizer                        & Adam              \\
      Scheduler                        & ReduceLROnPlateau \\
      Scheduler patience               & 20                \\
      Schedule factor                  & 0.7               \\
      Epochs (full field)              & 300               \\
      Epochs (stress-strain curve)     & 500               \\
      Training dataset                 & 80\%              \\
      Validation dataset               & 20\%              \\
      \hline
    \end{tabular}
    \label{tab:not_train_hyper}
  \end{table}

  \subsection{Model performance}
\begin{figure}[hb]
    \centering
    \includegraphics[width=\textwidth]{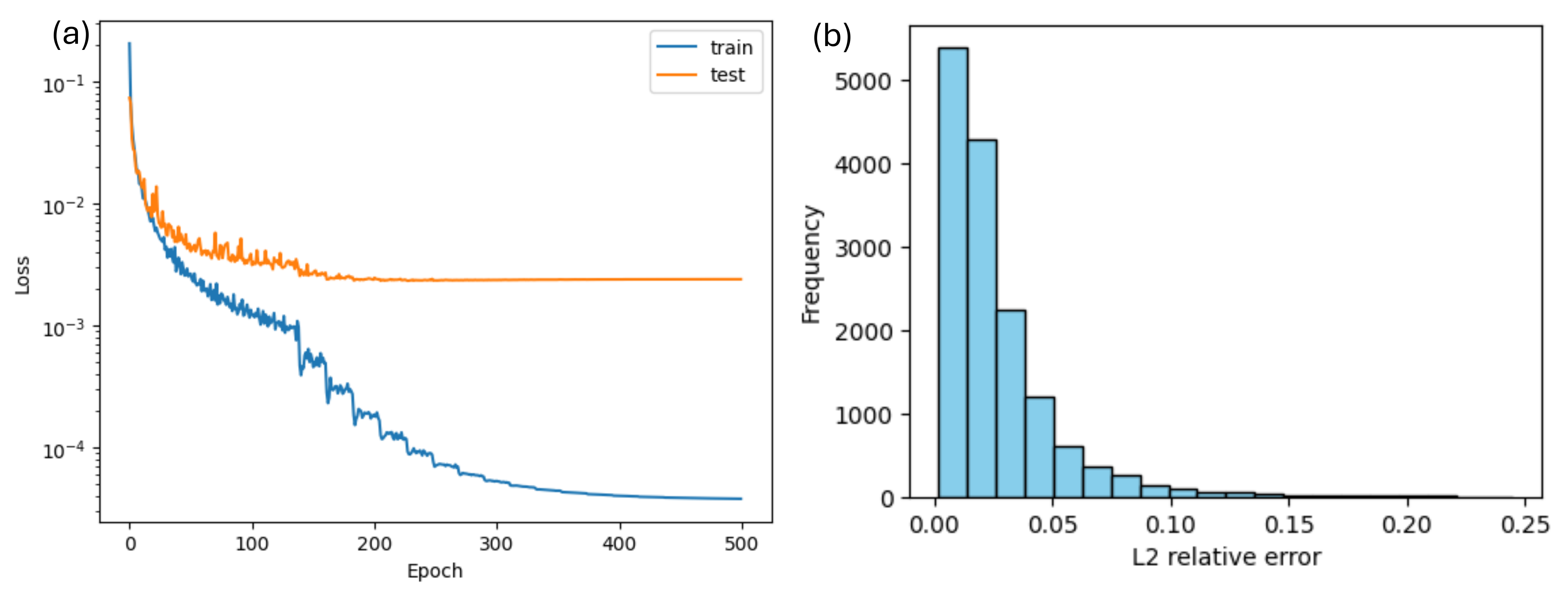}
    \caption{Performance of the NOT forward model for predicting the stress-strain
    curve from SDF. (a) shows the MSE loss during training, and (b) presents the
    $L_{2}$ relative error distribution for the test data. The mean $L_{2}$ relative
    error is 2.6\% with a standard deviation of 2.4\%.}
    \label{fig:S_notss_perform}
  \end{figure}

\begin{figure}[hb]
    \centering
    \includegraphics[width=\textwidth]{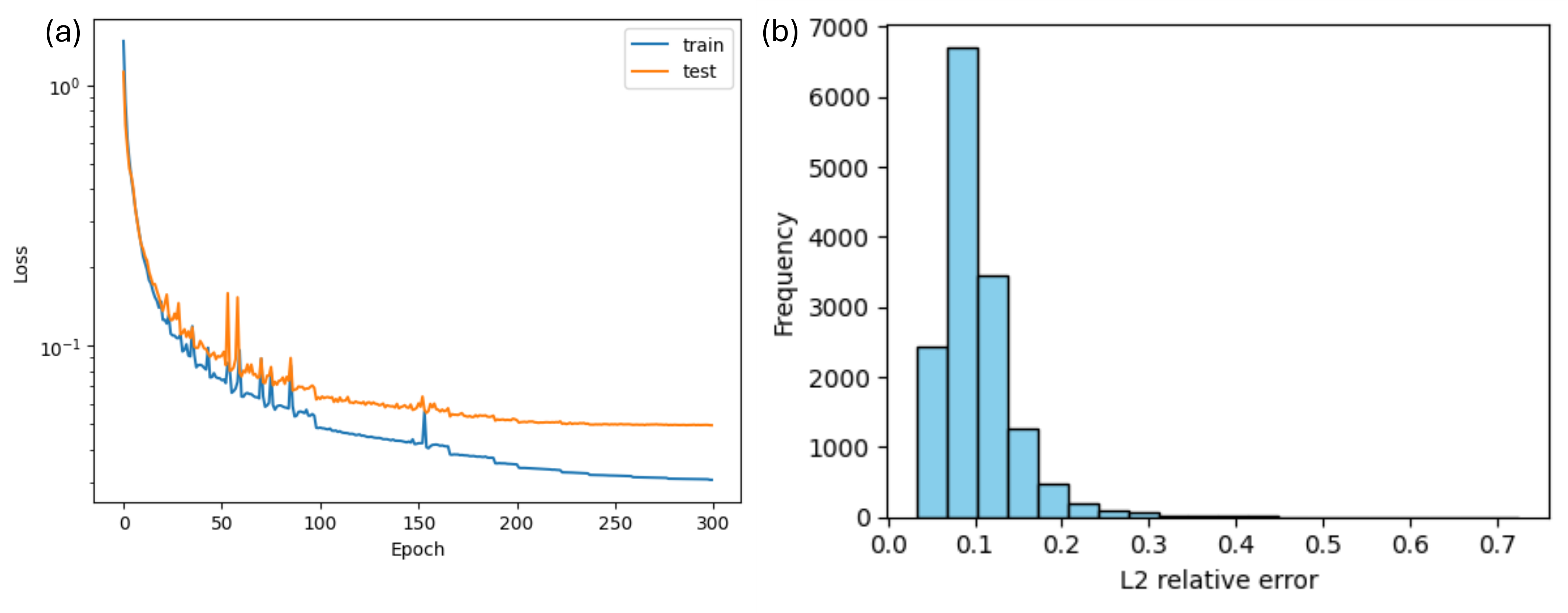}
    \caption{Performance of the NOT forward model for predicting Mises stress
    and displacement from arbitrary geometries. (a) shows the MSE loss during training,
    and (b) presents the $L_{2}$ relative error distribution for the test data
    of Mises stress and displacement. The mean $L_{2}$ relative error is 10.3\%
    with a standard deviation of 4.6\%.}
    \label{fig:S_notsu_perform}
  \end{figure}

\begin{figure}[hb]
    \centering
    \includegraphics[width=\textwidth]{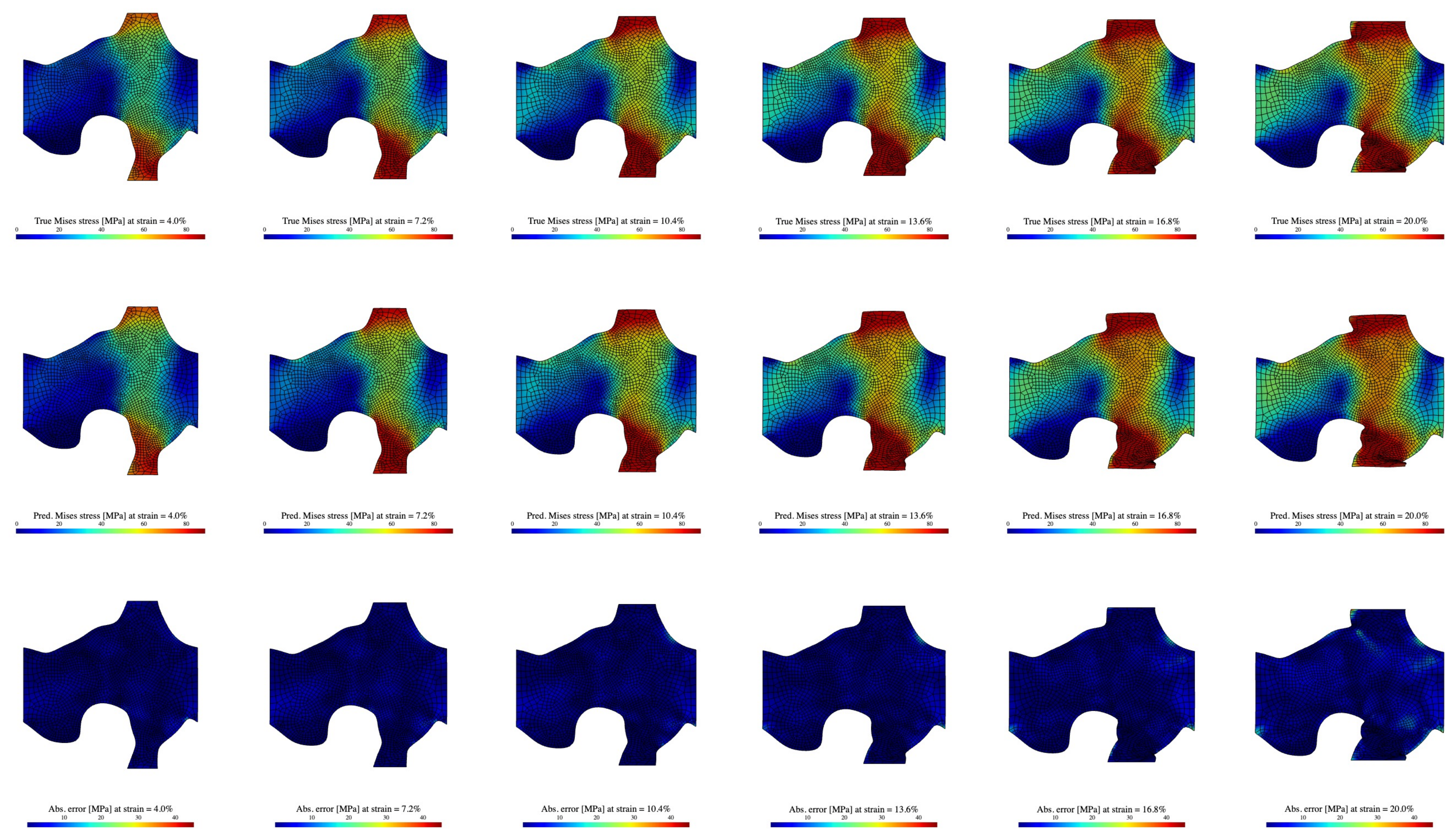}
    \caption{Comparison between the Mises stress and displacement field
    predictions with the corresponding FE ground truth at different strain
    $\varepsilon$ steps for the \textbf{best case} of the \textbf{test data}. The
    first row shows the true Mises stress under the true deformed shape, the
    second row shows the predicted Mises stress under the predicted deformed
    shape, and the third row shows the absolute error of the Mises stress under the
    true deformed shape. The $L_{2}$ relative error over the total 26 strain
    steps is 3.9\%.}
    \label{fig:S_notsu_perform_best}
  \end{figure}

\begin{figure}[hb]
    \centering
    \includegraphics[width=\textwidth]{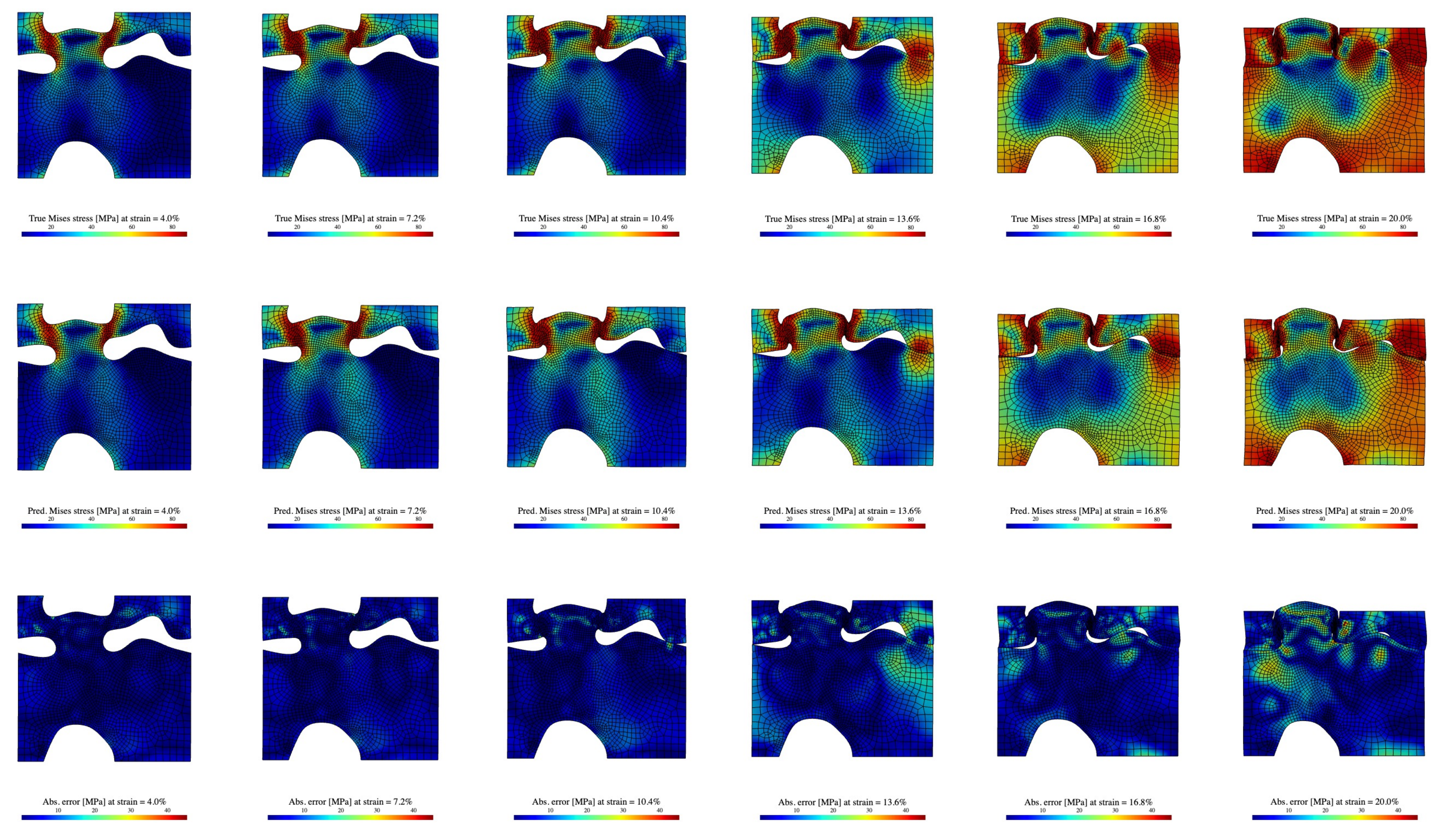}
    \caption{Comparison between the Mises stress and displacement field
    predictions with the corresponding FE ground truth at different strain
    $\varepsilon$ steps for the \textbf{worst case} of the \textbf{test data}. The
    first row shows the true Mises stress under the true deformed shape, the
    second row shows the predicted Mises stress under the predicted deformed
    shape, and the third row shows the absolute error of the Mises stress under the
    true deformed shape. The $L_{2}$ relative error over the total 26 strain
    steps is 72.6\%.}
    \label{fig:S_notsu_perform_worst}
  \end{figure}

  The model performance of the forward NOT model for predicting the stress-strain
  curve is shown in \cref{fig:S_notss_perform}, where (a) displays the MSE loss
  of the training history and (b) presents the $L_{2}$ relative error
  distribution of the test data. Despite the overfitting indicated in \cref{fig:S_notss_perform}(a),
  the $L_{2}$ relative error of the test data is as low as 2.6\% with a standard
  deviation of 2.4\%, demonstrating that the model generalizes well to the test
  data.
  The performance of the forward NOT model for predicting the solution fields is
  presented in \cref{fig:S_notsu_perform}. \cref{fig:S_notsu_perform}(a) shows the
  MSE loss during training, while \cref{fig:S_notsu_perform}(b) displays the $L_{2}$
  relative error distribution for the test data. The mean $L_{2}$ relative error
  is 10.3\% with a standard deviation of 4.6\%, indicating that the model generalizes
  well to the test data. The best and worst prediction cases from the test data
  are illustrated in \cref{fig:S_notsu_perform_best} and
  \cref{fig:S_notsu_perform_worst}, respectively. The $L_{2}$ relative error
  over the total 26 strain steps is 3.9\% for the best case and 72.6\% for the
  worst case.

  \clearpage

  \section{Inverse model}
  \label{s_sec:inverse}
  \subsection{Diffusion model theory}
  \label{s_sec:diffusion} In this work, the classifier-free guided diffusion
  model \citep{s_ho2022classifier} is trained to design the micro-structure of the
  unit cell of a metamaterial that can achieve a target stress-strain curve. The
  labeled data $y$ consists of the stress-strain curve obtained from the ABAQUS
  simulation of the corresponding random geometry. The corresponding SDF image serves
  as the design variable $\mathbf{x_0}$ for the diffusion model. By applying the
  marching algorithm on the designed SDF images to extract the contour with a
  value equal to 0, the boundary of the designed geometry is obtained. The model
  architecture is shown in \cref{fig:models}(d). For completeness, the denoising
  diffusion probabilistic model \cite{s_ho2020denoising} and the classifier-free
  guidance method \citep{s_ho2022classifier} are briefly reviewed here.

  Given an initial sample $\mathbf{x_0}$ from a prior data distribution
  $\mathbf{x_0}\sim q(\mathbf{x})$, the forward diffusion process of the denoising
  diffusion probabilistic model incrementally adds a small amount of Gaussian noise
  to the sample across $T$ steps. This creates a sequence of samples
  $\mathbf{x}_{1}, \mathbf{x}_{2}, \ldots, \mathbf{x}_{T}$. Each step in this diffusion
  process is regulated by a variance schedule $\left \{ \beta_{t}\in (0,1)\right
  \}_{t=1}^{T}$,
  \begin{subequations}
    \label{eq:s_diffusing}
    \begin{align}
       & q\left(\mathbf{x}_{t}|\mathbf{x}_{t-1}\right)=\mathcal{N}\left(\mathbf{x}_{t}; \sqrt{1-\beta_{t}}\mathbf{x}_{t-1}, \beta_{t}\mathbf{I}\right), \\
       & q\left(\mathbf{x}_{1: T}|\mathbf{x}_{0}\right)=\prod_{t=1}^{T}q\left(\mathbf{x}_{t}|\mathbf{x}_{t-1}\right).
    \end{align}
  \end{subequations}
  The sample $\mathbf{x}_{0}$ gradually loses its features and eventually
  becomes an isotropic Gaussian distribution for large $T$. This diffusion process
  has the elegant property that we can sample $\mathbf{x}_{t}$ at any time step $t$
  using the reparametrization trick,
  \begin{subequations}
    \label{eq:s_sampling}
    \begin{align}
       & \mathbf{x}_{t}= \sqrt{\bar{\alpha}_{t}}\mathbf{x}_{0}+ \sqrt{1 - \bar{\alpha}_{t}}\epsilon_{t},                                                                            \\
       & q\left(\mathbf{x}_{t}|\mathbf{x}_{0}\right) = \mathcal{N}\left(\mathbf{x}_{t}; \sqrt{\bar{\alpha}_{t}}\mathbf{x}_{0}, \left(1 - \bar{\alpha}_{t}\right) \mathbf{I}\right),
    \end{align}
  \end{subequations}
  where $\alpha_{t}= 1 - \beta_{t}$,$\bar{\alpha}_{t}= \prod_{i=1}^{t}\alpha_{i}$,
  and $\epsilon_{t}\sim \mathcal{N}(0, \mathbf{I})$. By reversing the diffusion
  process, we can sample from $q(\mathbf{x}_{t-1}| \mathbf{x}_{t}, \mathbf{x}_{0}
  )$ and ultimately generate the true sample $\mathbf{x}_{0}$ from Gaussian noise
  $\mathbf{x}_{T}\sim \mathcal{N}(0, \mathbf{I})$. Using Bayesian rule, the reverse
  posterior distribution conditioned on $\mathbf{x}_{0}$ is traced,
  \begin{equation}
    q\left(\mathbf{x}_{t-1}|\mathbf{x}_{t}, \mathbf{x}_{0}\right) = \mathcal{N}\left
    (\mathbf{x}_{t-1}; \tilde{\boldsymbol{\mu}}_{t}\left(\mathbf{x}_{t}, \mathbf{x}
    _{0}\right), \tilde{\beta}_{t}\mathbf{I}\right),
  \end{equation}
  where
  $\tilde{\boldsymbol{\mu}}_{t}\left(\mathbf{x}_{t}, \mathbf{x}_{0}\right) = \frac{1}{\sqrt{\alpha_{t}}}
  \left(\mathbf{x}_{t}- \frac{1 - \alpha_{t}}{\sqrt{1 - \bar{\alpha}_{t}}}\epsilon
  _{t}\right)$
  and
  $\tilde{\beta}_{t}= \frac{1 - \bar{\alpha}_{t-1}}{1 - \bar{\alpha}_{t}}\beta_{t}$.
  In a real generative process, we cannot directly evaluate
  $q\left(\mathbf{x}_{t-1}|\mathbf{x}_{t}, \mathbf{x}_{0}\right)$ because it
  requires the training sample and the entire diffusion noise dataset. Instead,
  we propose a posterior distribution
  $p_{\theta}\left(\mathbf{x}_{t-1}|\mathbf{x}_{t}\right)$ to approximate the
  true posterior distribution
  $q \left (\mathbf{x}_{t-1}|\mathbf{x}_{t}, \mathbf{x}_{0}\right )$,
  \begin{equation}
    \label{eq:s_approx_posterior}p_{\theta}\left(\mathbf{x}_{t-1}|\mathbf{x}_{t}\right
    ) = \mathcal{N}\left (\mathbf{x}_{t-1}; \boldsymbol{\mu}_{\theta}\left (\mathbf{x}
    _{t}, t\right ), \mathbf{\Sigma}_{\theta}\left(\mathbf{x}_{t}, t\right)\right
    ).
  \end{equation}
  The reverse denoising process is then controlled by
  \begin{equation}
    \label{eq:reverse}p_{\theta}\left(\mathbf{x}_{0:T}\right) = p\left(\mathbf{x}
    _{T}\right) \prod_{t=1}^{T}p_{\theta}\left(\mathbf{x}_{t-1}|\mathbf{x}_{t}\right
    ),
  \end{equation}
  where $\mathbf{\Sigma}_{\theta}\left(\mathbf{x}_{t}, t\right)$ is the same as $\tilde
  {\beta}_{t}\mathbf{I}$, and
  $\boldsymbol{\mu}_{\theta}\left(\mathbf{x}_{t}, t\right )$ has the same form
  as $\tilde{\boldsymbol{\mu}}_{t}\left(\mathbf{x}_{t}, \mathbf{x}_{0}\right)$
  but with learnable parameters $\theta$,
  \begin{equation}
    \label{eq:s_approx_mu}\boldsymbol{\mu}_{\theta}\left(\mathbf{x}_{t}, t\right
    ) = \frac{1}{\sqrt{\alpha_{t}}}\left(\mathbf{x}_{t}- \frac{1 - \alpha_{t}}{\sqrt{1
    - \bar{\alpha}_{t}}}\epsilon_{\theta}\left(\mathbf{x}_{t}, t\right)\right ).
  \end{equation}
  The optimization objective of the denoising diffusion probabilistic model is to
  maximize the log-likelihood $\log p_{\theta}\left(\mathbf{x}_{0:T}\right)$,
  which results in the following loss function:

  \begin{equation}
    L = \left|\epsilon_{\theta}\left(\mathbf{x}_{t}, t\right) - \epsilon_{t}\left
    (\mathbf{x}_{t}, t\right)\right|.
  \end{equation}

  Training a NN $\epsilon_{\theta}\left(\mathbf{x}_{t}, t\right)$ and minimizing
  the loss function $L$ leads to $\boldsymbol{\mu}_{\theta}\left(\mathbf{x}_{t},
  t\right ) \approx \tilde{\boldsymbol{\mu}}_{t}\left(\mathbf{x}_{t}, \mathbf{x}_{0}
  \right )$, thus making the proposed posterior distribution
  $p_{\theta}\left(\mathbf{x}_{t-1}|\mathbf{x}_{t}\right)$ close to the true
  posterior distribution
  $q\left(\mathbf{x}_{t-1}|\mathbf{x}_{t}, \mathbf{x}_{0}\right)$. The above reverse
  generative process is random and not controlled by any specific target. In
  specific design tasks, we aim to generate fields that represent the target,
  such as the force-displacement curve, which requires training the NN $\epsilon_{\theta}
  \left (\mathbf{x}_{t}, t \right)$ with conditional information. To incorporate
  the condition information $\mathbf{y}$ into the diffusion process, \citet{ho2022classifier}
  proposed a classifier-free guidance method that incorporates the scores from a
  conditional diffusion model $p_{\theta}(\mathbf{x}|\mathbf{c})$ and an
  unconditional diffusion model $p_{\theta}(\mathbf{x})$. The noise estimators $\epsilon
  _{\theta}\left(\mathbf{x}_{t}, t\right )$ of $p_{\theta}(\mathbf{x})$ and $\epsilon
  _{\theta}\left(\mathbf{x}_{t}, t, \mathbf{c}=\mathbf{y}\right)$ of $p_{\theta}(
  \mathbf{x}| \mathbf{c})$ are trained in a single NN $\epsilon_{\theta}\left(\mathbf{x}
  _{t}, t, \mathbf{c}\right )$. Here, $\epsilon_{\theta}\left (\mathbf{x}_{t}, t,
  \mathbf{c}=\mathbf{y}\right )$ is trained with paired data $( \mathbf{x}, \mathbf{c}
  =\mathbf{y})$, and $\epsilon_{\theta}\left (\mathbf{x}_{t}, t\right)$ is
  trained with data $\mathbf{x}$ only, i.e.,$( \mathbf{x}, \mathbf{c}= \emptyset
  )$. During the training process, the condition information $\mathbf{c}$ is
  randomly set as $\mathbf{c}=\emptyset$ or $\mathbf{c}= \mathbf{y}$ for sample
  $( \mathbf{x}, \mathbf{y})$ at different time steps $t$, allowing the model to
  learn to generate fields both conditionally and unconditionally. During the
  reverse inference process, $\epsilon_{\theta}$ in \cref{eq:s_approx_mu} is replaced
  by the linear summation of conditional and unconditional noise estimators,
  \begin{equation}
    \label{eq:s_guidance}\epsilon_{\theta}\left(\mathbf{x}_{t}, t, \mathbf{c}\right
    ) = (1 + w)\epsilon_{\theta}\left(\mathbf{x}_{t}, t, \mathbf{c}=\mathbf{y}\right
    ) - w \epsilon_{\theta}\left(\mathbf{x}_{t}, t, \mathbf{c}=\emptyset \right),
  \end{equation}
  where $w \ge 0$ is the guidance weight.

  \subsection{Model architectures}
  The classifier-free guided diffusion model is designed to predict the SDF of
  the unit cell of metamaterial that can achieve a target stress-strain curve. The
  model architecture is summarized in \cref{tab:invse_Unet}. It consists of
  three main components: a stress-strain (SS) curve encoder, a time encoder, and
  a U-Net encoder-decoder. The outputs of the SS curve encoder and the time
  encoder are fed into the residual blocks in the U-Net.
  \begin{table}[!h]
    \centering
    \caption{Geometry encoder architecture of the forward NOT model (B is batch size).}
    \begin{tabular}{|c|c|l|l|}
      \hline
      Network       & \textbf{Layer \#} & \textbf{Layer type (Description)}                            & \textbf{Output shape}            \\
      \hline
      Encoding      & 1                 & Input (SS curve)                                             & (B, 51)                          \\
      stress-strain & 2                 & MLP with SiLU                                                & (B, 32)                          \\
                    &                   & (channels: 51$\rightarrow 32 \rightarrow 32 \rightarrow 32$) &                                  \\
                    & 3                 & Linear (Output)                                              & (B, 16)                          \\
      \hline
      Encoding      & 1                 & Input (time t)                                               & (B, )                            \\
      time t        & 2                 & Time embedding+Linear (Output)                               & (B, 64)                          \\
      \hline
      U-Net         & 1                 & Input (SDF at time t)                                        & (B, 1,120,120)                   \\
      encoder       & 2                 & Conv2D                                                       & (B, 16, 120, 120)                \\
                    & 3                 & Residual Conv Block                                          & (B, 16, 120, 120)                \\
                    & 4                 & Conv2d (DownSample)                                          & (B, 16, 60, 60)                  \\
                    & 5                 & Residual Conv Block                                          & (B, 32, 60, 60)                  \\
                    & 6                 & Conv2d (DownSample)                                          & (B, 32, 30, 30)                  \\
                    & 7                 & Residual Conv Block                                          & (B, 64, 30, 30)                  \\
                    & 8                 & Conv2d (DownSample)                                          & (B, 64, 15, 15)                  \\
                    & 9                 & Residual Conv Block + self-attention                         & (B, 128, 15, 15)                 \\
      \hline
      U-Net         & 10                & Residual Conv Block                                          & (B, 128, 15, 15)                 \\
      bottle-neck   &                   & + self-attention + Residual Conv Block                       &                                  \\
      \hline
      U-Net         & 11                & Concatenate                                                  & (B, 256, 15, 15)                 \\
      decoder       & 12                & Residual Conv Block + self-attention                         & (B, 128, 15, 15)                 \\
                    & 13                & Conv2D (UpSample)                                            & (B, 128, 30, 30)                 \\
                    & 14                & Concatenate                                                  & (B, 192, 30, 30)                 \\
                    & 15                & Residual Conv Block + self-attention                         & (B, 64, 30, 30)                  \\
                    & 16                & Conv2D (UpSample)                                            & (B, 64, 60, 60)                  \\
                    & 17                & Concatenate                                                  & (B, 96, 60, 60)                  \\
                    & 18                & Residual Conv Block                                          & (B, 32, 60, 60)                  \\
                    & 19                & Conv2D (UpSample)                                            & (B, 32, 120, 120)                \\
                    & 20                & Concatenate                                                  & (B, 48, 120, 120)                \\
                    & 21                & Residual Conv Block                                          & (B, 16, 120, 120)                \\
                    & 22                & GroupNorm+SiLU+Conv2D (Output)                               & (B, 1, 120, 120)                 \\
      \hline
                    & 1                 & Inputs                                                       &                                  \\
      Residual      &                   & (image, encoded SS curve, encoded t)                         & (B, in, W, H), (B,16), (B,64)    \\
      Conv Block    & 2                 & Conv2D,SiLU+Linear, SiLU+Linear                              & (B, out, W, H), (B,out), (B,out) \\
                    & 3                 & --,Expand+repeat,Expand+repeat                               & (B, out, W, H)$\times3$          \\
                    & 4                 & Concatenate                                                  & (B, 3*out, W, H)                 \\
                    & 5                 & Conv2d                                                       & (B, out, W, H)                   \\
                    & 6                 & Residual sumation                                            & (B, out, W,H)                    \\
      \hline
    \end{tabular}
    \label{tab:invse_Unet}
  \end{table}

  The overview of the training hyperparameters of the classifier-free guided
  diffusion model is shown in \cref{tab:invse_train_hyper}.
  \begin{table}[!h]
    \centering
    \caption{Training hyperparameters of the classifier-free guided diffusion
    model.}
    \begin{tabular}{c|c}
      \hline
      \textbf{Hyperparameter} & \textbf{Value}    \\
      \hline
      Batch size              & 64                \\
      Initial learning rate   & 1e-3              \\
      Optimizer               & Adam              \\
      Scheduler               & ReduceLROnPlateau \\
      Scheduler patience      & 20                \\
      Schedule factor         & 0.7               \\
      Epochs                  & 300               \\
      Training dataset        & 80\%              \\
      Testing dataset         & 20\%              \\
      Time steps ($T$)        & 500               \\
      \hline
    \end{tabular}
    \label{tab:invse_train_hyper}
  \end{table}

  \newpage
  \subsection{Model performance}
\begin{figure}[!h]
    \centering
    \includegraphics[width=4in]{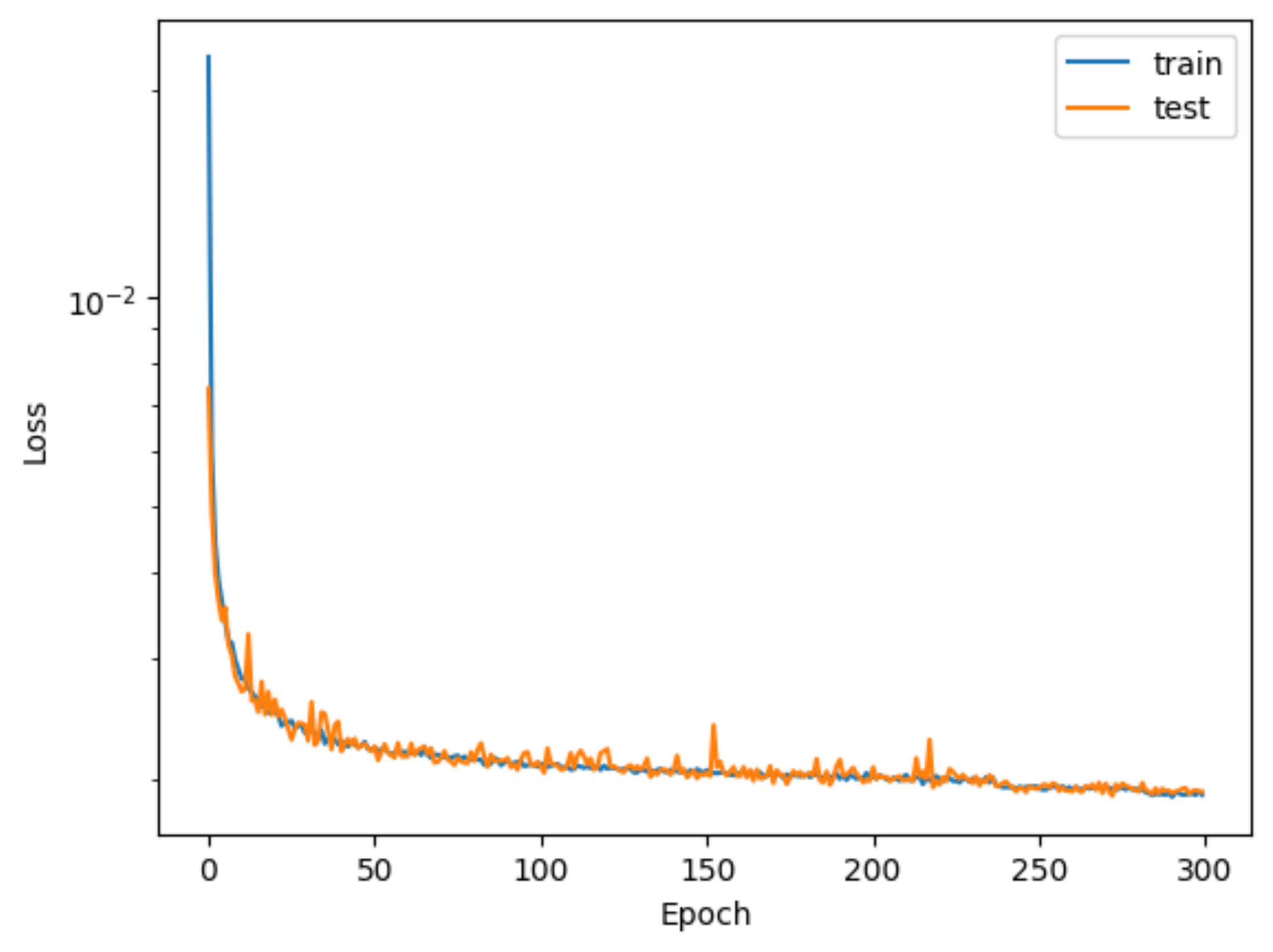}
    \caption{MSE loss of training history of the diffusion model.}
    \label{fig:S_inv_loss}
  \end{figure}

  The model performance of the classifier-free guided diffusion model is shown
  in \cref{fig:S_inv_loss}

  \newpage
  \section{Computational efficiency}
  \label{s_sec:efficiency}

  The FE simulations were executed on CPUs of the Delta machine at the National
  Center for Supercomputing Applications (NCSA) at the University of Illinois at
  Urbana-Champaign. Model training was conducted on a single NVIDIA H100 GPU on the
  DeltaAI machine at NCSA. Inference tasks for the forward NOT model and the
  inverse design model were performed on an NVIDIA A100 GPU on the Delta machine.
  The computational efficiency of data generation, model training, and inference
  is summarized in \cref{tab:efficiency}.

  \begin{table}[!h]
    \centering
    \caption{Computational efficiency of data generation, model training, and
    inference.}
    \begin{tabular}{c|c}
      \hline
      \textbf{Task}              & \textbf{Time}     \\
      \hline
      FE simulation (1 cpu)      & 2.3 min/sample    \\
      NOT training (SS curve)    & 16 s/epoch        \\
      NOT training (full field)  & 96 s/epoch        \\
      Diffusion model training   & 47 s/epoch        \\
      NOT inference (SS curve)   & 1.6e-4 s/solution \\
      NOT inference (full field) & 1.0e-2 s/solution \\
      Inverse design             & 0.46 s/solution   \\
      \hline
    \end{tabular}
    \label{tab:efficiency}
  \end{table}

\end{document}